# Pearlnecklacelike chain conformation of hydrophobic polyelectrolyte: a SANS study of partially sulfonated polystyrene in water.

*M.N. Spiteri[1], C.E. Williams[2], F. Boué[1\*]*

1 Laboratoire Léon Brillouin, CEA/CNRS UMR12 CE Saclay, 91191 Gif-sur-Yvette Cedex, France

fboue@cea.fr

2 Laboratoire des Fluides Organisés, UMR 7125, CNRS/Collège de France, 11 place Marcelin Berthelot, 75005 Paris, France.

*This paper is dedicated to Claudine Williams.*

ABSTRACT The form factor of partially sulfonated polystyrene PSSNa (degree of sulfonation $f = 1$, 0.72, 0.64 and 0.36), at polymer concentration 0.17M and 0.34M, without or with added salt (0 M, 0.34M, & 0.68M), is measured by Small Angle Neutron Scattering using the Zero Average Contrast method. The total scattering function is also measured, allowing us to extract the distinct interchain function and an apparent structure factor. The main result is the behavior of the form factor which shows contributions of spherical entities as well as extended chain parts. This is striking for 0.64, while for $f = 0.36$ the sphere contribution is more dominant. The conformation does not depend on polymer concentration. When salt is added, the sphere sizes do not vary, but the contribution attributed to the stretched parts does vary very much like for fully sulfonated PSSNa. Discussion of the interchain contribution permits to establish the level of interpenetration: solutions are interpenetrated for f= 0.64, and at the limit for f=0.36. The theoretical pearl necklace model appears very suitable to modeling the results. Comparisons are made with analytical calculation and simulation data of pearl necklace. While the respective roles of Rayleigh transition, heterogeneous architecture, and strong hydrophobicity of non sulfonated PS monomers remain under discussion, the data give an accurate three dimensional image of the pearl necklace in solution.

# 1 Introduction.

Random sulfonation of the common synthetic polymer polystyrene (PS) followed by neutralization results in a copolymer poly-((styrene)$_{1-f}$-*co*-(styrene sodium sulfonate)$_f$), abbreviated PSSNa (or PSS). This is a very versatile species, with strikingly different properties according to the degree of sulfonation, i.e. to the linear charge fraction of the chain *f*.

At low *f* (less than about 15%), the polymer can be ***dissolved in organic solvents*** of the PS sequences, and shows unique thickening and gelling properties; in the dry state it has the mechanical properties of a physical network. Numerous experimental and theoretical studies have shown that, due to the low permittivity $\varepsilon$ of solvent molecules and styrene monomers, attractive dipolar interactions between non-dissociated ions pairs induce self-organization, resulting in numerous applications of these so-called "*ionomers*" [1].

For larger values of f, typically f > 0.30, the copolymer becomes ***water soluble***. Each styrene sulfonate unit is dissociated into a positively charged chain unit and a negatively charged counterion (Na$^+$). In particular when all monomers are sulfonated, at $f$ =1, PSSNa aqueous solutions of this polymer have been shown to have all the characteristic properties of a polyelectrolyte solution. In the attempt to validate the predictions of scaling theories [2-5], PSSNa solutions have been considered in particular as a model for structural studies in semi-dilute regime [6-11]. The conformation shows no sign of temperature dependence, suggesting good hydrophilicity [8]. Tensioactive properties are only detected below f = 0.9.

For intermediate charge fractions, typically 0.30< *f* <0.9, there is a possibility that the short range hydrophobic attraction between uncharged monomers competes with the strong long range electrostatic repulsion between the charged species to determine the chain conformation and the structure of the solution. In contradistinction to the first two cases, i.e. ionomers and fully charged polyelectrolyte, which are well documented and understood, this range of intermediate charge fractions had first received little attention until a few years. It is all the more surprising since many polyelectrolytes of practical interest result from a competition between hydrophobic parts and ionic groups which have been attached to make the chain water-soluble. It is these "hydrophobic polyelectrolytes» that will be the general concern of this paper, beyond the polymer studied here partially sulfonated polystyrene.

Though partially charged, the chains considered here have more than 35% of charged monomers: under these conditions, as long as a chain has a string-like conformation, the electrostatic interactions between charged monomers are larger than $k_BT$, which is called the ***strong coupling limit***. This leads to the prediction of ***counterions condensation***: the electrostatic potential on the highly charged chain is so high that some counterions remain in its close vicinity. According to the Manning - Oosawa model [12], the counterions will condense until the average distance between charged units is equal to the Bjerrum length $l_B$ = 7 Å, the distance over which the electrostatic energy equals $k_BT$. In practice for PSSNa the charge to charge distance before condensation is a ~ 2.5 Å, which will thus, after condensation, bring the effective charge fraction down to $f_{Manning}$ = $a/l_B$~ 0.35. Therefore, for 0.35 <f<1, the effective charge $f_{eff}$ of a polyion in a string-like conformation should be constant. This prediction has proved to be qualitatively correct for most *hydrophilic*, charged, single chain. More subtle situations, like coexistence of monovalent and divalent counterions which induces on fully sulfonated PSSNa [13] a succession of condensation situation depending on their relative ratio, confirm very nicely Manning's predictions.

For hydrophobic polyelectrolytes, the situation is very different. Theories have been erroneous in overlooking hydrophobic interactions under the misleading assumption that, in the strong coupling limit polyelectrolyte properties were entirely dominated by electrostatics. Experiments on PSS have revealed a large variation of $f_{eff}$ with f, in the range [0.35, 1] and a strong reduction with respect to Manning's expected values except for f=1 [14-17]. Similarly the structure of the semi-dilute PSS solutions, id est the interchain correlations as seen by scattering [14-15], was also strongly dependent on f and different from what was observed for a hydrophilic polyelectrolyte [6]. Clearly these scattering experiments showed that the hydrophobic effects have a profound influence on the properties of PSSNa. A proposed interpretation was that chains are contracted by hydrophobic effects, and fluorescence measurements suggested hydrophobic domains.

At the same time have appeared models for chain conformation and solution structure: whereas, in analogy with a chain in bad solvent [2, 18] simple transition between extended chain and collapsed chain, isotropic or a cigar-shaped succession of blobs (Khokhlov, [19]), was first proposed, Kardar and Kantor [20, 21] and Dobrynin, Obukov, Rubinstein [22-24] have applied the concept of Rayleigh instability [25] to charged polymers in poor solvent, leading to a pearl necklace chain conformation, in parallel with Solis-de la Cruz[26], Lyulin et al [27], and others [28]. The chain is now a succession of extended parts (strings) and compact collapsed parts (pearls); most of the chain segment mass is concentrated in the pearls, with included or condensed counterions. Strings and pearls coexist along the chain in a dynamic way. Numerical simulations have also shown the existence of pearls [29-34], including the annealed case (charges rearrange dynamically along the chain like for a weak polyacid for example), for which the phase diagram also contains a direct first-order transition towards a collapsed chain [35]. In the quenched case, such models could apply to PSSNa at f<1. Other observations have been reported recently for polymers deposited onto a surface: ellipsometry [36] permits to access to a thickness which depends on rate of charge f, while pearl-like objects are visible by AFM, on systems such as poly(2-vinylpyrridine) and poly(methacryloyloxyethyl dimethylbenzylammonium chloride) [37] and polyvinylamine [38]. More recently an AFM investigation under different conditions of controlled adsorption (mica as well as lipid membranes) has been conducted on partially sulfonated PSSNa [39].

Obviously, scattering techniques can be useful on these systems, by yielding the form factor P(q) of the chain in solution. It is possible to extract P(q) using SAXS or SANS in dilute enough solutions. However, semi-dilute solutions are often encountered in practice for polyelectrolytes, where chains are often extended. Moreover poly-S-co-SNa has the impressive property that such solutions remain stable at high concentration; there is no phase separation. In such conditions, SANS combined with Zero Average Contrast (see explanations below) is a unique tool to provide direct observation of the form factor in semi-dilute regime. Such observations were published sometimes ago in a Ph. D work [9]. Meanwhile, several small angle scattering papers providing indirect evidence for pearl necklace shape have preceded the present paper. A first approach [40] was on a collapse induced via changing the solvent quality by adding acetone to water. A second approach was slightly different: it deals with the impact of specifically interacting alkaline earth cations which neutralise anionic chains via complex bond formation with the anionic residuals [41, 42]. Along both strategies, SANS and SAXS was performed under dilute solution conditions. In view of the recent large amount of experimental, theoretical, and computational work on these systems, data on form factor in semi-dilute solutions(which high concentration also should provide a better accuracy), in the line of [9], appear useful. We note that, beyond former work on PSSNa by X-ray scattering [14-17], ellipsometry [36], and more recent work like reflectivity [45], narrow comparisons with AFM [39] are now possible since the solutions of PSSNa were also prepared under the same conditions.

The aim of this paper is thus to report direct form factor SANS observations of PSSNa chain conformation for 0.35< f <1 in semi-dilute solutions. In such regime of concentration, the chains interact and are much interpenetrated. Former scattering measurements were done on PSSNa solutions where all chains where labeled with respect to the solvent, using Small Angle Neutron Scattering (SANS [6]) as well as X rays (SAXS, [14-15]). The obtained quantity is called "total scattering"; it displays a maximum, the well known "polyelectrolyte peak" in absence of salt. "Total" scattering does not differentiate the correlations between units pertaining to two different chains from correlations between units of the same chain. To measure the conformation of one chain, we need to access the second ones (inside the same chain), even in a semi-dilute solution where chains are highly interpenetrated. Besides formerly used an extrapolation method [7], a more sophisticated scattering technique has been used, the so-called Zero Average Contrast method (ZAC). This has already given access to conformation of fully sulfonated PSSNa chains (f = 1) in semi dilute solution [8-10]; the conformation has been described by the wormlike chain model, with a persistence length $l_p$. Now, for f <1, it is interesting to check how this wormlike chain is modified by hydrophobic interactions, which we show below.

## 2 Experimental details.

### 2.1 Polymers.

#### 2.1.1 Synthesis.

All polyelectrolytes investigated here have been obtained by post-sulfonation of polystyrene, in the laboratory. Perdeuterated polystyrene (d-PS) as well as non-deuterated polystyrene (h-PS), with very close degree of polymerization (d-PS, $N_{wD}$ =652, h-PS, $N_{wH}$ = 625) and a narrow mass distribution (see Table 1) were purchased from Polymer Standard Service (Mainz, Germany). The three rates of charge studied are very close for the two deuterated and the non-deuterated chains: f ~ 1(see Table 1), f = 0.64 +/- 0.01 and f = 0.36 +/- 0.02 (an extra sample with f = 0.72 has also been studied). The first one, f = 1, has been synthesized using the Vink method, while the two others have been synthesized using the Makowski method.

The Vink method [43] is commonly used to reach total sulfonation; one starts from a polystyrene solution in cyclohexane (a PS theta solvent at 35°C), which is poured onto a mixture of sulfuric acid with phosphoric acid. After stirring for half an hour, the mixture is let to rest for decantation. Separation in three phases is triggered by addition of ice. The phase containing PSSH (polysulfonic acid) is extracted and neutralized by an excess of Sodium hydroxide. The obtained PSSNa solution is dialyzed against deionized water until the conductivity of the external dialysis bath remains stable. The solutions are then concentrated with a rotating evaporator and finally freeze-dried. The resulting white powder is better stored away from light.

The Makowski method [44] has been used for partial sulfonation, after advices of W. Essafi [14]. As the Vink one, it is a phase transfer, interfacial, reaction. A dichloroethane PS solution is mixed with acetic acid and sulfuric acid in proportions depending on the desired rate of charge. A white layer appears between the two media. After 30 to 60 min, the aqueous phase is neutralized with Sodium hydroxide, dialyzed, concentrated and ice dried.

**2.1.2 Characterization**.

The molecular weight of parent PS chains has been measured by Size Exclusion Chromatography (see Tables 1 and 2). The molecular weight distribution has been measured by Size Exclusion Chromatography (in water) for f=1 and is given in Table 1; the molecular weight and the polydispersity have been slightly increased, may be due to a slight bridging, but the chromatogram curves are very close to the PS parent ones. For f<1, we had no standard for a SEC characterization. Baigl et al [45, 46] have shown that the distribution of the degree of polymerization is negligibly different from the one of parent PS for not too high DP (like here), and also that Makowski sulfonation is very reproducible. The charge fraction f has been determined by elementary analysis (ratio between sodium and carbon in the dry sample). We see in Table 2 that f is the same for the non deuterated chain and the deuterated chain. We see here from these measurements that reproducibility applies for d-PS. This is crucial for ZAC measurements, of course.

No measurement of the sulfonated units distribution along the chain was performed on the polymers used in the present study. Recent careful checks [47] using a fluorescence ray specific to sulfonated units have reinforced the picture of a random poly (styrene-co-styrene sodium sulfonate) copolymer.

The aqueous solutions were prepared by dissolving the desired amount of polymer in the solvent. Solvent was $D_2O$ (purchased from Eurisotop and used as received) and chains were non deuterated PSS (h-PSS) for the case where all chains are labeled with respect to solvent. Solvent was a mixture of deionized $H_2O$ (resistivity 18 MΩ), and pure $D_2O$ for ZAC measurements (see below). The $H_2O/D_2O$ fractions suitable to the ZAC method were calculated for each chemical composition, id est for each f (assuming an average monomer density of scattering length for both styrene and styrene sulfonate units). We would like to note that viscosity was always low, even at the largest concentrations studied here. In water, even when highly salted in the case where salt was added up to 3M NaCl, PSSNa f=1 is dissolved after 12 to 24 Hours. When f <1 it is better to allow for more time for dissolution (here 48 hours).

We took care of the water content of dry PSSNa. For f = 1 it is admitted that at least 10% water is still present; for <1, we checked (Karl Fisher tests done at CNRS Vernaison) that 10 to 14% weight could have been adsorbed. This corresponds to more than one water molecule per charged unit on average.

All solutions have been filtered on 0.22 μm Millipore membranes except those with f = 0.36, for which it turned out to be impossible; since filtration may modify too much the polymer concentration. The final polymer concentration was measured after filtration by titration of the carbon in the sample (COT, total organic carbon, Dohrmann); accuracy is 5%. Separate checks were done for H and D samples.

*2.2 SANS measurements.*

**2.2.1 The Zero average contrast method.**

Let us recall the fundamentals of the most convenient method by which the form factor of a chain among others can be obtained. We start from the general expression of the scattered intensity:

$$I(q) \text{ (cm}^{-1}\text{, or Å}^{-1}) = (1/V) \cdot d\Sigma/d\omega = I(\vec{q}) = \frac{1}{V}\left\langle \sum_{i,j} k_i k_j \exp(i\vec{q}(\vec{r}_i - \vec{r}_j)) \right\rangle \quad (1a)$$

where $V_{mol\,i}$ et $V_{mol\,s}$ are respectively the partial molar volumes of the repeating unit i and the solvent s, and where $k_i$ (cm or Å) = $b_i - b_s \cdot (V_{mol\,i}/V_{mol\,s})$ is the "contrast length" between one repeating unit of scattering length $b_i$ and molar volume $V_{mol\,i}$, and a solvent molecule ($b_s$, $V_{mol\,s}$).

Assume first that all chains are labeled with respect to solvent; here we dissolve H-PS into $D_2O$. The concentration is $c_p$, in mole/L (or mole/ Å$^3$), so the total volume fraction of chains is $\Phi_T = N_{Av} \cdot c_p \cdot V_{mol\,i}$, where $N_{Av}$ is the Avogadro number. Then for all i, we have $k_i = k_H$ (the value is given in Table 3), and

$$I(q) \text{ (cm}^{-1}\text{, or Å}^{-1}) = (1/V) \cdot d\Sigma/d\omega = k_H^2 \, S_T(q) \quad (1b).$$

Using Å and Å$^{-1}$ as the units for $k_H$ and I(q), we obtain $S_T(q)$ in Å$^{-3}$. Quite generally,

$$S_T(q) = S_1(q) + S_2(q), \qquad \text{Å}^{-3} \qquad (2a),$$

where

$$S_1(q) \,(\text{Å}^{-3}) = \frac{1}{V}\left\langle \sum_{\substack{\alpha \text{ avec}\\ \beta=\alpha}} \sum_{i,j} \exp(i\vec{q}(\vec{r}_i^{\,\alpha} - \vec{r}_j^{\,\beta})) \right\rangle \qquad (2b)$$

corresponds to the correlations between monomers i,j of the same chain $\alpha = \beta$ (intrachain scattering) and

$$S_2(q) \,(\text{Å}^{-3}) = \frac{1}{V}\left\langle \sum_{\substack{\alpha,\\ \beta\neq\alpha}} \sum_{i,j} \exp(i\vec{q}(\vec{r}_i^{\,\alpha} - \vec{r}_j^{\,\beta})) \right\rangle \qquad (2c)$$

corresponds to the correlations between monomers i,j of two different chains $\alpha$ and $\beta \neq \alpha$ (interchain scattering).

Assume now that *only a fraction of the chains are labeled*. We use a mixture of a number fraction $x_D$ of d-PSS chains ($k_i = k_D$) and $(1-x_D)$ of h-PSS chains ($k_i = k_H$). The total volume fraction of chains in the solution is the sum of the volume fractions of the two types of chain, $\Phi_T = \Phi_H + \Phi_D$ (we have in general $V_{molH} = V_{molD}$, so $\Phi_D/\Phi_T = x_D$ and the equation $\Phi_T = N_{Av}.\, c_p.\, V_{mol\,H}$ is still valid, $c_p$ being the total polymer molar concentration). The scattered intensity (1a) becomes:

$$I(q)\,(cm^{-1}) = (1/V).\, d\Sigma/d\omega = \{[(1-x_D)\,k_H^2 + x_D\,k_D^2]\,S_1(q)\} + \{[(1-x_D)\,k_H + x_D\,k_D]^2\,S_2(q)\} \quad (3).$$

This second type of labeling allows us to suppress the interchain contribution $S_2(q)$, if we can have $(1-x_D)\,k_H + x_D\,k_D = 0$.

This is possible if we use as a solvent a mixture of $H_2O$ and $D_2O$: then the average scattering length of the solvent $b_S$ can be varied. In the equation above, the symmetric case $k_H = -\,k_D$ (which also implies $x_D = 0.5$) is the most efficient situation in term of intensity. This is obtained if $b_S/V_S$ is made equal to the arithmetic average of $b_H/V_{mol\,H}$ and its pendent $b_D/V_{mol\,D}$. For PSS, this corresponds to a solvent constituted of 71% $H_2O$ and 29% $D_2O$ [8-10]. We write $|k_{ZAC}| = -\,k_H = k_D$, and Eq. (3) gives:

$$I(q) = k_{ZAC}^2\,S_1(q) \quad (4),$$

which permits a direct measurement of intrachain scattering of one chain among the others, even in the semi-dilute regime. The different contrast length values are listed in Table 3. The values evaluated for the contrast lengths of the Na counterions with the $H_2O/D_2O$ mixture used here are low; their contribution to the scattering have therefore been neglected. This has been confirmed by a more refined evaluation accounting for hydration [48, 49]. The $S_1(q)$ limit at q tending to zero is

$$\lim_{q \to 0} S_1(q) = c_p.\, N_{Av}\, N_w\,,$$

where $c_p$ should be expressed in mole/ $Å^{-3}$. Hence, from the definition of the form factor, we can write

$S_1(q) = c_p\, N_{Av}\, N_w\, P(q)$.

To give an order of magnitude, the zero q limit of $S_1(q)$ is close to 0.2 $Å^{-3}$ for $c_p$= 0.34 M. This corresponds for I(q) to about 10 $cm^{-1}$. Values obtained for $N_w$ (Table 4) will be discussed below.

The ZAC technique has been used since on polyelectrolytes by other authors [49-51].

**2.2.2 Measurements and data treatment.**

SANS measurements have been performed on the PACE spectrometer at the Orphée reactor of LLB, CEA- Saclay, France (www-llb.cea.fr). A range of scattering vector $q = (4\pi/\lambda)\sin\theta/2$ between $5.10^{-3}$ and 0.4 $Å^{-1}$ was covered using the following two settings: D=0.92m, λ=5Å and D=3.02m, λ= 12.5Å. Samples were contained in 2 mm thick quartz cells. All measurements were done at room temperature.

All data have been normalized using the incoherent scattering of a high proton content sample, here light water; the latter has been calibrated to obtain absolute values of (1/V). $d\Sigma/d\omega_{water}$ in $cm^{-1}$, using Cotton's method [52]. The solvent contribution is subtracted from these corrected data. The subtraction of the solvent incoherent background is however quite delicate and deserves further remarks. At large q (> 0.2 $Å^{-1}$) especially, the coherent part of the intensity is very small compared to the background due to incoherent scattering, which has several origins:

- incoherent scattering from $H_2O$ and $D_2O$ in the solvent

- hydration water molecules adsorbed on the polymer dry chains (more than 10% in weight, see above)

- protons from the h-PSSNa, and deuterons from d-PSSNa .

- protons from water vapor molecules after contact with air.

These small contributions are delicate to estimate and thus make us unable to know the exact quantity to subtract with accuracy better than 3%.

Such uncertainty has little influence for small q but can lead to different shapes for large q.

Also, mixing the components leads to an extra flat scattering, called Laue scattering or sometimes "mixing incoherent", which is actually the coherent scattering from the mixture of small elementary components such as different molecules in a solvent. For best results, and to eliminate as much as possible effects of multiple scattering (though they are here very weak) which involve the cell geometry, we have prepared under the same conditions some blanks, by mixing $H_2O$ with $D_2O$, aiming at the same flat intensity; it was also checked that they had the same neutron transmission.

## 3 Scattering results.

### 3.1 Total scattering: comparison with previous data.

Before examining the single chain contribution $S_1(q)$, let us have a look at $S_T(q)$, id est all pair correlations when all chains are identically labeled (in this case h-PSS in $D_2O$), in order to ensure that these polymers show the general behavior obtained previously by Essafi et al (the fact that our chains are shorter and more monodisperse is not relevant here). Indeed we find the same features [14, 15]:

(i) without added salt $S_T(q)$ (Fig. 1) is characterized by a maximum, as for a fully charged PSSNa, but depending upon f:

- the peak abscissa q* decreases strongly when f is decreased (Fig. 2), meaning that the characteristic distance between chains increases. The height of the maximum also increases strongly when f is decreased, by more than a factor 10 when passing from f = 1 to f = 0.36.

- for each f, $q^* \sim c_p^\alpha$, where $\alpha$, which equals 0.5 for f=1, decreases slightly at lower f: for f = 0.34, $\alpha < 0.4$.

(ii) upon addition of salt (NaCl, Note 1), for f = 0.64 as well as f = 0.36 , the "polyelectrolyte peak" vanishes (Fig. 3) . The zero q limit increases noticeably for lower f, suggesting larger elementary objects.

(iii) finally, when tending towards low q, the curve upturns, an effect which strengthens at low f, suggesting in all cases large scale inhomogeneities [53-56]. Lower qs are required to conclude whether they strengthen at lower f.

Let us recall that if only electrostatics were involved, Manning condensation should, for all f > $f_{Manning}$ = $l_B$/ a, bring f down to $f_{Manning}$ = $l_B$/ a= 0.36. Therefore the scattering maximum should not depend on f (let us remind that we are not concerned here by the case f < $f_{Manning}$, for which theory introduces the electrostatic blob, and predicts q* ~ $f^{1/3}$ as checked for hydrophilic partially charged polyions like AMPS [14-15] and poly(Acrylic Acid-co- Acrylamide) [58-60 ]).

At variance with Manning's theory, the variation with f is strong: Fig. 2 shows that q* ~ $f^{0.89 \pm 0.08}$ for $c_p$ = 0.34M and q* ~ $f^{0.81 \pm 0.11}$ for $c_p$ = 0.17 M. It is worth recalling at this stage that Essafi, Lafuma and Williams have found for f = 0.36 at low concentration ($10^{-4}$ M) that the structure factor is close to the one of small charged spheres, the spheres charge being that measured by osmotic pressure. So the system seems to pass from interpenetrated extended chains to more compact chains further from each other. This explains the strong variation of q* with f.

The vanishing of the peak is observed for any f; it is complete when $c_s$ = $c_p$. This agrees with electrostatic repulsion vanishing when the distance between species worths several times the screening length.

*3.2 Chain form factor (ZAC conditions)*

The case f=1 has been studied in detail formerly [7-10]. It will be in this paper the basis for a comparison with the case f<1.

**3.2.1 Effect of degree of sulfonation.**

The scattering profiles for the 3 degrees of sulfonation f at a constant total polymer concentration $c_p$ = 0.34 M under ZAC conditions are shown on Fig.4 in a $q^2 S_1(q)/c_p$ representation (Kratky plot). The three curves are remarkably different. Previous detailed investigations of the fully charged chains, f=1, have shown that it can be described by a wormlike chain [9-10]. None of its characteristics are seen for the two lower charge fractions. Especially striking is the appearance of a broad "bump" in the q range whereas the curve for f=1 displays an horizontal inflexion, a short so-called "Gaussian plateau" ($q^2 S_1(q)$ ~ cst, since $S_1(q)$ ~ 1/$q^2$ at q <1/$l_p$, $l_p$ is the persistence length). The possible "remains" of this phenomenon are still visible under the form of a short plateau around q = 0.05 Å$^{-1}$ for f = 0.64. More precisely we can distinguish three zones.

*In the small q range*, $q^2 S_1(q)$ is first increasing, before that the plateau appears for f=1. This range corresponds to the Guinier regime:

$c_p/S_1(q) = c_p k_H^2/I(q)$

$= (1/N_{Av} \cdot N_w) \cdot (1+ q^2 <R_g^2>_z/3), \qquad qR_g < 1 \quad (5)$

where $N_w$ is the weight average number of monomer per chain, and $R_g = <R_g^2>_z^{1/2}$ (Å) the radius of gyration after z averaging over the mass distribution (of both H and D chains, which are very close). A Zimm representation $c_p/S_1(q)$ versus $q^2$ (not shown here), allows one to extract the radius of gyration $R_g$ as the slope of the straight line, and $N_w$ from its extrapolation to zero q. The values for the three f's are noted in Table 4. For f = 0.64 the value is close to the one for f=1. For f=0.36, on the contrary $N_w$ is larger (1130). It is important to note that the determination of the two parameters $N_w$ and $R_g$ is very sensitive to the existence of large scale heterogeneities [53-57]. In principle ZAC should make them invisible but upon any slight unaccuracy or uncomplete mixing, they contribute to the intensity I(q) in this q range because they are very large, and $R_g$ and $N_w$ may be overestimated. For f=0.36, large aggregates may be present because this solution could not be filtered (see Experimental section). However, in spite of a high apparent $N_w$, the measured radius for f=0.36 is noticeably smaller.

From these values of $R_g$, we see that the polyions have an increasingly compact conformation when f is decreased. But we are still far from a complete collapse of the chain: had the chain expelled all the solvent, the radius of gyration would be:

$$R_{gcoll} = \sqrt{(3/5)} \, (3/4\pi) \, (M_w/d \, N_{Av})^{1/3} \sim 20 \, \text{Å} \qquad (6),$$

taking $d = 1.96 \, 10^{-24} \, \text{g/Å}^3$ for the density of PSSNa. The ratio $R_g/R_{gcoll}$ is given in Table 5. It is also revealing to compare with the radius of gyration of a PSSNa chain (f = 1) of same $M_w$ measured in presence of 3M of added salt, for which all electrostatic repulsion is suppressed ("neutral chain"). The measured $R_g$ is 100 Å [9-10]. We see that this value of $R_g$ is reached here for 0.64, i.e. that the trend to contraction is already balancing the electrostatic repulsion for the global size of the polyion. For f = 0.36, at the same global scale, the chain is already contracted compared to a "neutral chain", by a factor 3 in global volume.

*In the medium q range,* $q^2 S_1(q)$ keeps increasing for f≠1, until a maximum is reached at $q_{bump}$, and then decreases. This decrease corresponds to $S_1(q) \sim q^{-\alpha}$, with $\alpha > 2$. Such a maximum, and its subsequent decrease, are signatures of scattering objects more compact than a Gaussian chain. One textbook example is a Gaussian ring (id est a Gaussian chain with its ends kept together). In that case it is known that $q^2S_1(q)$ displays a maximum of abscissa close to 1/ $R_g$. In our case the maximum abscissa is not compatible with the value of $R_g$ measured in the Guinier range. Another limit case is that of a sphere. Fig. 5 shows an attempt to fit our curves to the form factor of a sphere, in the q range of the $q^2S_1(q)$ maximum: we find that the maxima correspond to radii of R = 16 ± 1 Å for f = 0.64 and R = 24 ± 1 Å for f= 0.36. These radii R give radii of gyration ($R_g = \sqrt{(3/5)}$ R) much lower than the values we measured in the Guinier regime.

***In the high q range,*** $q^2S_1(q)$ is increasing again. Although we must remain careful about background subtraction error in this region, the features are nevertheless clear. For f = 0.64, the curve joins the one for f = 1. The variation is therefore close to $S_1(q) \sim q^{-1}$, as for f = 1: this suggests a picture of a chain with wormlike parts. This $q^{-1}$ behavior is hardly visible for f = 0.36, but it may be masked by the tail of the strong decrease observed in the second range.

To summarize, at this stage the analysis of single chain scattering profile, in a $q^2S_1(q)$ representation, has shown a compaction of the chain ($R_g$) as f decreases. It has also hinted the presence of compact dense objects whose size is smaller than Rg and increases with f. Meanwhile, the contribution of some flexible parts is visible for f=0.64. We are thus tempted to imagine a composite conformation in between a wormlike chain and some small compact parts.

***Log-log plot.*** We also show a comparison in log-log plot, in Fig. 6; it gives a better insight of the low q regime (it does not privilege large qs like the $q^2S_1(q)$ plot).. We visualize successively, for low sulfonation rates (here f = 0.64 and also for another set of data recorded with f=0.72), the low q Guinier regime, then a continuous decrease close to a power law, on which is superimposed a supplementary contribution in the range [0.04 Å$^{-1}$, 0.25 Å$^{-1}$]; this contribution is the pendant of the oscillation which appears in the $q^2S_1(q)$ plot.

**3.2.2 Variation with polyion concentration and salt concentration.**

***Polymer concentration***. If we now compare the form factors for two values of $c_p$ = 0.34 M and 0.17M, we observe a surprising result: the form factors ***hardly vary with $c_p$*** for the partially charged PSS at f = 0. 64 as well as for f = 0.36.

If we first look at the radii of gyration (Table 4) we see that concentration effects are lower and lower when decreasing f. Whereas, for f = 1, $R_g$ passes from 171Å to 197 Å when $c_p$ passes from 0.34 M to 0.17 M, for f = 0.64, it only changes from 97 Å to 105 Å, the variation being within the limit of experimental error. Finally, for f=0.36 the apparent variation (within the same experimental error) is even reversed.

If we now look at the $q^2S_1(q)$ plots of Fig. 7, the lack of change is striking:

- for f = 0.64, at intermediate q, the maxima are superposed; the **dense spheres do not change at all**. At lower q and larger q, differences are visible. They are close to what we would have in these q ranges for f = 1, namely the $q^2S_1(q)$ plateau and the foot of the increasing part at large q would both be shifted upward at larger $c_p$. For f = 1 it is a consequence of a decrease of the persistence length when $c_p$ increases. Therefore if we imagine that

$S_1(q) = A(c_p, c_s) S^{spheres}(q, c_p, c_s) + B(c_p, c_s) S^{strings}(q, c_p, c_s)$ (7),

(Note that equation (7) is oversimplified because it omits cross-correlations between the strings and pearls). Data suggest that A and B do not vary much, that $S^{spheres}(q)$ does not vary and that $S^{strings}$ vary like for f=1.

- for f=0.36, the sphere contribution is also insensitive to $c_p$. The string contribution is very minute, if it even exists. It seems to be larger at lower concentration.

In summary the conformation is dominated by the spheres contribution, which is insensitive to polymer concentration.

*Salt concentration* The effect of added salt has been investigated for f = 0.64 only. The radius of gyration decreases when salt concentration $c_s$ increases from 0 M (97 Å) to 0.34M (76 Å) and 0.68M (66 Å). This can be due to retraction of the stretched part, or to shrinking of the dense parts.

If we now look at the $q^2S_1(q)$ plots of Fig. 8, we see that the abscissa maximum in the medium q range(around q = 0.15 Å$^{-1}$) and its height hardly change. This suggests that: compact parts are insensitive to co-ions in the same way they are to counterions and polyion concentration. In the frame of equation (7), the term $A(c_p, c_s)$ $S^{spheres}(q)$ stays unchanged. The variation of $R_g$ can thus only come from the other parts contribution. Indeed in Fig.8 at low q, the height of the short flat horizontal part just before the bump (in other words the horizontal inflexion at q = 0.05 Å$^{-1}$ ), which is visible in the three curves clearly increases with $c_s$. The height increase factor is about 2 when passing from $c_s$ =0 M to $c_s$ = 0.68 M. Let us assume that this "short plateau" corresponds to the contribution of the wormlike part of chains; then its height would be proportional to 1/ $l_p$, $l_p$ being the persistence length. If the chains considered here contain strings blocks, we can thus interpret the increase of the plateau by a decrease of the persistence length inside these strings, due to electrostatic screening. This is supported by the fact that the factor 2 is close to the change in $l_p$ observed for f=1 [9-10] for the same salt content (Note 2). In the frame of equation (7), the term $S^{strings}(q)$ varies like for f=1. This reasoning assumes that the variation would not come from the factor B, which we cannot prove. However, a change in B (string parts collapsing into spheres) should induce a change in the mass or size of the spheres, which is not seen.

Returning to the decrease of $R_g$, let us note that if the dense parts are unchanged, the decrease could be due to a change in the string parts. In such case, the contribution of the latter to $R_g$ should vary like $l_p^{0.5}$. This is close to what we find for the global $R_g$, within the error (rather large here).

*Log-log plot*. We can also visualize these effects more obviously on the log-log plots of Fig. 9. The variation of $S_1(q)$ corresponding to the "short plateau" is seen here on a wider scale.

Between 0.02 and 0.05 Å$^{-1}$ the straight part of the curves is shifted towards higher values when salt is added. At larger q (> 0.005 Å$^{-1}$), they join, whereas the shoulder visible between 0.05 Å$^{-1}$ and 0.25 Å$^{-1}$ for low salt content is masked when $c_s$ reaches 0.68 M. In the latter case, one sees the advantage of the $q^2S_1(q)$ plot, for which the corresponding contribution ("bump") is still apparent even at $c_s$ = 0.68 M.

*3.3 Interchain correlations.*

When looking at the maximum in $S_T(q)$, one is tempted to interpret such so-called "polyelectrolyte peak" as the maximum of a structure factor. However, $S_T(q)$ contains the contribution of $S_1(q)$. If the chains were fully disinterpenetrated (and centrosymmetric in average), the ratio $S_T(q)/S_1(q)$ would be equal to the S(q). Let us consider here that $S_T(q)/S_1(q)$ is an apparent structure factor, $S_{app}(q)$, and see what information can be extracted. $S_{app}(q)$ becomes closer to a real structure factor when chains become more and more collapsed.

At this point, let us also recall that Essafi et al [14] considered a $q^2S_T(q)$ plot. Had $S_1(q)$ varied like $1/q^2$, such a plot would have been an approximation of $S_T(q)/S_1(q)$. The $q^2S_T(q)$ plot displayed a maximum, or a "shoulder". As evoked above, a numerical simulation (L. Belloni) for a solution of charged spheres well described $q^2S_T(q)$ for a PSSNa (f=0.36) solution at much lower concentration, where one is sure to be in dilute regime ($c_p$ < 10$^{-2}$ M).

The apparent structure factor plotted in Fig. 10 displays a maximum which is shifted to lower q when f is decreased, like for $S_T(q)$. Within the error bars, small oscillations can be noticed, as some characteristic of a stronger order than for f = 1. In order to confirm this impression, we show in Fig.

11 the distinct pair scattering $S_2(q)$ which displays more marked oscillations. The first maximum abscissa for $S_2(q)$ is the same as for $S_T(q)/S_1(q)$. Its value, $q_{maxS}$, is always larger than $q^*$, and its variation with f is $q_{maxS} \sim f^{0.63\pm0.03}$ as shown in Fig.12, which is softer than the one of $q^* \sim f^{0.85}$.

The $S_T(q)/S_1(q)$ ratio remains lower than 1, at variance with a classical structure factor, which should tend to 1 at large q (this is better obeyed for f = 0.36). The same trend was observed for all fully charged PSSNa solutions [9]. This is probably due to the fact that $S_T(q)$ and $S_1(q)$ have not been measured in the same solvent (i.e. not in the same mixture of $H_2O$ and $D_2O$). It has been shown that measurements of $S_T(q)$ for non deuterated PSSNa in the same solvent as for the ZAC measurements make $S_{app}(q)$ tend to 1 at large q [9].

Similar plots are obtained for $c_p$ =0.17M, with maxima slightly more peaked (this effect of $c_p$ was also observed for f = 1[9]).

***Interchain distance.*** Let us now discuss the value of the abscissa, $q_{maxS}$ for the maximum in $S_{app}(q)$. More rigorously we must consider the maximum in $S_2(q)$, the maximum abscissa of which are anyway very close.

The question is whether this maximum can be interpreted by a mean distance between parts of the chain or between the chains themselves. The answer depends upon the degree of interpenetration. For f = 1, the case has been discussed earlier. The chains are stretched by electrostatic repulsion, the chains are very interpenetrated and the maximum in $S_{app}(q)$, $S_T(q)$ or better $S_2(q)$, can be interpreted by a ***mean distance between chain strands***, $\xi \sim 2\pi/q^*$. For weaker f, $q^*$ is lower: this means that the interchain distance increases. We also know from $S_1(q)$ that the global size of the chains decreases. These two effects both lower the degree of interpenetration. To estimate this degree, we compare the global chain size $R_g$ with the radius $R_{overlap}$ of the spherical volume available per chain (in other words the volume occupied by one chain if $c_p$ was equal to the overlapping concentration $c_p^*$). $R_{overlap}$ is defined such as

$$c_p^* = (N / N_{Av}) / (4\pi R_{overlap}^3 / 3) / \alpha) \qquad (8)$$

($c_p$ is in mole/L; we take a compaction factor $\alpha = 0.74$). Table 7 shows that the chains are still interpenetrated ($R_g > R_{overlap}$) for f = 0.64, whereas for f = 0.36 there is a transition towards the dilute regime. The conclusion is the same when replacing $R_g$ by the half end-to-end distance assuming a Gaussian conformation, or the radius of a sphere of same $R_g$ (see caption of Table 7).

We can also compare chain sizes with some estimate of the distance between chains, extracted from the maximum abscissa: $2\pi / q_{maxS}$ or $2\pi / q^*$ (Table 8). These values are smaller than the overlapping distance between two charged spheres, equal to 2 . $R_{overlap}$ (given in Table 7). This is particularly marked if we privilege the $q_{maxS}$ values. We again conclude that chains are interpenetrated for f=0.64 and not completely disinterpenetrated for f=0.36.

When adding salt for f = 0.64, we had observed the vanishing of the maximum in $S_T(q)$. On the contrary, $S_{app}(q)$ (Fig. 13) still shows a maximum (as for f = 1, not shown here). But the oscillations vanish when salt is added. The same is seen on the plot of $S_2(q)$ in Fig.14. The maximum is better due to a correlation hole than to a strong repulsion. Similar behaviors are seen in neutral polymer solutions.

## 4 Discussion.

### 4.1 Conformation models: success of the pearl necklace model.

The form factor profile, in particular for f=0.64, displays some characteristics of a collapsed chain, and some of a wormlike chain at large q. Several possible situations can be imagined.

*Mixture of two kinds of chains*. First one could imagine that a fraction of the chains would be wormlike (WLC), meanwhile other chains would be dense globules. The intrachain signal would be a simple addition of the two types of conformations. This coexistence could come from a heterogeneous charge distribution – which we do not expect (see Section 2.1), or from a more physical origin, like a first-order transition predicted by theoretical models, formerly, or recently by simulations [35]. But such a "mixture picture" is in contradiction with the fact that for f=0.64 because the medium q range is well described by the form factor of a sphere of radius 16 Å: this value is clearly lower than the one which we can calculate for a completely collapsed chain, $(5/3)^{1/2} R_g \sim 26$ Å. A mixture would be possible for f = 0.34, since the medium q range gives R = 24 Å which is close to 26 Å. However, the total scattering $S_T(q)$ shows always a unique well defined maximum, which is also in contradiction with the "mixture picture".

*First models of composite chains.* Second, let us then look for model of composite chains. First, we tried to model our data by the form factor of a Gaussian chain made of freely jointed $N_p$ pearls of diameter D, containing $n_p$ monomers [61, 62]. Here the pearls are adjacent, they are the repeating units; hence we have a simple product:

$$P(q) = P_{SD}(q;D/2) \cdot P_G(q; N_p, D) \qquad (9a)$$

with: $\qquad P_{SD}(q;D/2) = \left[\frac{3}{u^3}(\sin u - u\cos u)\right]^2 , \; u = q\,D/2 \qquad (9b)$

and $\qquad P_G(q; N_p, D) = \frac{2}{x^2}\left[\exp(-x) + x - 1\right], \; x = q^2 \cdot R_g^2 \qquad (9c)$

with, for our Gaussian coil, $R_g = \left(\frac{NpD^2}{6}\right)^{1/2} \qquad (9d)$

The best fit is shown in Fig. 15 for f = 0.36. Even if we can fit at low q, the fit at large q is obviously poor: the maximum in the calculated model is very broad, due to the Gaussian nature of the model, in as much as the number of pearls found is large ($N_p = 149$; this makes also the pearls diameter rather small, 12 Å). For f = 0.64, it is impossible to obtain a high enough abscissa for the maximum in $q^2S_1(q)$, and the required spheres radius becomes extremely small.

We thus need a conformation where **both the spherical conformation and the stretched conformation** at large q are accounted for. A first possibility could be a chain which is locally stretched, and globally collapsed. But this is in contradiction with the maximum in $q^2S_1(q)$, which is well pronounced and characteristic of much smaller spheres. The other possibility is the coexistence of pearls and strings on the same chain, which we discuss now.

*Pearl necklace model.* The composite conformation of the pearl necklace model, introduced above, appears to be a fundamentally better model. This is due to its intrinsic structure of pearls separated by some linear strands, or strings, together with a flexible conformation at large scale. In principle, under such conditions, the signal is a sum of the two types of objects plus a cross-correlation term. At low q, it is governed by a global radius of gyration, which can have the right value to agree with data. In the medium q range, the signal is often dominated by the spheres signal, where most of the chain mass is often concentrated. The string contribution, and the cross-terms, can be neglected. At large q, the spheres signal decreases as $q^{-4}$; it can therefore be neglected, whereas the signal of the strings, which decreases more slowly, in $q^{-1}$ only, can be seen. The cross-correlation will be negligible both because of the $q^{-4}$ variation of the spheres form factor, and because large qs correspond to scales shorter than the sizes of the spheres and strings, which then appear uncorrelated at this scale. The fact that the pearl size increases with decreasing f is predicted by the Dobrynin - Rubinstein model.

This model **agrees** in particular **with the effect of salt** on $S_1(q)$, detailed in section 3.2.2. As partially discussed there, if the chain was made only of strings with a wormlike conformation, we would observe the following:

- in the range $1/R_g < q < 1/l_p$, the intensity varies as $(1/l_p) \cdot 1/q^2$, so $q^2S_1(q)$ displays a plateau of height $1/l_p$. Since $l_p$ decreases with $c_s$, the plateau level increases.

-         in the range $q > 1/l_p$, the intensity varies as $1/q$, independently of $l_p$, since we measure the scattering of the locally rodlike chain.

The first behaviour is observed on the "short plateau" in $q^2S_1(q)$, for $q \sim 0.05$ Å$^{-1}$, as shown in Fig. 8 (or equivalently in Fig. 9 on the straight part in log-log plot for q between 0.02 and 0.05 Å$^{-1}$). The second behavior is observed at high q ($1/l_p$) for $q > 0.25$ Å$^{-1}$. So the total intensity behaves as if it was the sum of a string contribution varying like in a wormlike chain plus a spherical contribution independent of salt concentration.

When the polymer concentration is changed from 0.34 M to 0.17 M for the same f = 0.64, the nominal ionic strength, due here to counterions, should decrease by a factor 2, and the "short plateau" decrease also by typically $2^{1/2}$. However the plots show only slight decrease changes even in the short plateau region. This may be due to the fact that the local counterion concentration inside the chain (polyions are not strongly interpenetrated) is not the nominal one. It can be influenced by condensation phenomena around the pearls.

An interpretation of these behaviours is also given by Liao et al, when commenting the results of their simulations which we report in the next section.

***Comparison with calculations and simulations***. In addition to the qualitative discussion given just above, we have the possibility of comparing our data to several available data obtained either by analytical or by numerical simulations. It is beyond the scope of this paper to make fits. But detailed graphic comparisons are given in Appendix. We summarize them here.

As said in Introduction, Micka, Holm and Kremer [30, 31], and later Limbach and Holm [32-34] have made numerical simulations showing clearly the existence of pearls on real space sketches. From these data, they have also calculated the form factor. Schweins and Huber [63] have calculated an analytical form factor of a chain made alternately of rods of constant length A separated by pearls of constant radius R, and compared their calculation to the results of Limbach and Holm [34]. Since in chains of common molecular weight, the simulations show that pearls are present in a small number, even as low as two or tree, Schweins et al have focussed their comarison on dumbbells and "trimpbells". The data are very marked by the fact that the distance A is constant, as could be predicted from the analogy with a Rayleigh transition. This produces an important oscillation with a local maximum in the log-log plots [57]. This oscillation is followed at larger q by a shoulder corresponding to the size of the pearls; this results in a kind of double oscillation, followed at larger q by a series of minima and maxima characteristic of the form factor of spheres. This succession of the two oscillations is depending upon the ratio between A and the pearl radius R, and its amplitude depends also upon the number ratio of monomers belonging to the pearl over monomers belonging to the strings. The model can fit some data from Limbach and Holm which display smoothed double oscillations. But comparisons of our data with both the analytical [63] and the simulated data[34] show in Fig. A.1 that such a privileged distance A is not clearly visible in our experiments.

More recently, Liao, Dobrynin and Rubinstein [64] ran numerical simulations of pearlnecklaces. These simulations do not show evidence of a privileged interpearl distance. The authors show actually fits of our data, in the case $f = 0.36$. We thus focused here on comparison for $f = 0.64$. Fig. A. 2 shows qualitatively that a proper fit is possible.

*In summary,* both a direct analysis and a comparison with numerical data show that the Dobrynin - Rubinstein picture of a sequence of strings and spheres ("pearls"), of increasing size when the rate of charge decreases, agrees quite well with the shape of the polyion form factor.

*4.2 Origin of pearls: Rayleigh transition or heterogeneous hydrophobic structure?*

Accepting that the conformation is a necklace of pearls, we have still to understand the origin of the pearls. In the theoretical picture leading to a pearl necklace structure, a kind of Rayleigh transition is proposed under the assumption that the chain has a homogeneous architecture. All repeating units ("monomers") are equivalent. The chain is immersed in a bad solvent, and has a given linear charge: both these properties are considered at a global level, id est are averaged over all segments. In practice, PSSNa in water does not precisely correspond to the polyelectrolyte considered by the model. It is a multiblock copolymer, poly-(SSNa-co-S) with a heterogeneous architecture. A memory of this could be kept when dissolved in water: for uncharged units water is an extremely bad solvent, while it is a good one for the charged units. One could then imagine a picture where some hydrophobic parts, coming from the same sequence, or from different several chains, would be localized within dense cores, as for surfactants micelles. The pearls could then correspond to these cores. Essafi et al have performed measurements of pyrene fluorescence which suggest the existence of regions able to trap pyrene, a hydrophobic non polar solvent [17], and this is also suggested by light scattering measurements [57]. The cores of PS sequences (or enriched in PS sequences) may localize pyrene (though one does not observe in pyrene fluorescence the kind of pronounced transition which is seen for micelles). Another point which could be in favor of this heterogeneous picture is the fact that the pearl size does not depend on the salt or polymer concentration (in the very limited range explored here).

However, several of these arguments can be contradicted. The fact that the distribution of the PS sequence is rather well statistical (random) is commonly accepted, from former studies up to recent studies [46]. The situation can then be more subtle, the solvent molecules undergoing some spatial reorganization averaged over hydrophobic and hydrophilic entities. We tried to detect by SANS the existence of well defined hydrophobic cores by using labeled organic solvent (toluene) mimicking pyrene, while the polymer signal was matched [65]: the observation was negative. The reasons for pearl size being insensitive to counterions and co-ions could be found in the high condensation expected around the pearls. Indeed, condensation effects are much stronger than predicted by Manning, as observed and reestablished recently by Essafi et al.[17]; they signal a very low fraction of effectively free charges, as low as $f_{eff}$= 0.04 for f = 0.36. The degree of condensation may be governed by the inner polarity of the pearls, as suggested by these authors [17]. However they conclude that this is not contradicting a pearl necklace conformation. Let us also recall that, although we are in semi-dilute regime, we have no sign of gelation, contrary to what observed for polyelectrolytes with hydrophobic parts. To be simple, we try to illustrate in Fig.16 a comparison between the pearl necklace model (right hand side) and the actual chain arrangement (left hand side).

## Conclusion

Measurements of the chain form factor of partially charged - partially hydrophobic PSSNa display a very systematic picture as a function of charge rate, concentration of polymer and added salt, which is consistent with the structure of interchain correlations. The chains are contracted when f decreases: their radius of gyration has decreased, though there is no complete collapse. For f = 0.64 at the least, wormlike chain characteristics are still visible: $q^{-1}$ variation at large q, "short plateau" at low q. The existence of clusters of small size, is strongly suggested (15 Å for f = 0.64 and 25 Å for f = 0.36). They are probably present within most of the chains. All this suggests a composite structure of strings and more compact pearls. Comparisons with analytical expressions and simulations of pearl necklace conformation are satisfactory; they do not explicitly show the presence of a well defined distance between pearls. The conformation does not vary when the polymer concentration is divided by 2. The dense clusters size does not vary either when salt is added.

Such insensitivity to salt could make us think that the pearls are aggregates of non sulfonated sequences existing along the chain (non random distribution). This also would explain their small size. However the organization of the solvent may average local effects. Also, no chain association is evidenced from the low viscosity of these solutions, although we are in a regime of weakly overlapped chains, as estimated from the values of radii of gyration and interchain distances. The pearl size may result from interactions large enough to be insensitive to the change in ionic strength and not permeable to the added salt because they are not polar.

One main aim of this paper is to make these data available. Obviously, detailed comparisons with theoretical predictions and simulations should be done.

***Acknowledgments***: We thank W. Essafi for her advices on partial sulfonation of our samples, and Patricia Lixon, Service de Chimie Moléculaire, CEA Saclay, for her great help in the characterization (COT, SEC). FB thanks Michel Rawiso for his support and suggestions. Claudine, it's so sad to finish the paper without you.

***Note 1***: We added salt only for the intermediate rate of charge (f = 0.64), in order to avoid reaching demixing for the lowest rate f = 0.36. However, a later study of Essafi shows that it is possible to add 0.3M of salt for $c_p$ = 0.34M, at the least.

***Note 2***: For totally charged PSSNa, at $c_p$ = 0.34 M, passing from $c_s$ = 0 to $c_s$ = 0.68 M corresponds to a variation of I = ½ (f . $c_p$ + 2 $c_s$) by a factor 5. Since for these solutions $l_p \sim I^{1/3}$, this gives a factor $5^{1/3} \sim 1.7$ for $l_p$. This is of the order of magnitude of the increase in height of the very short flat part (corresponding to the usual $q^2 S_1(q)$ plateau of a Gaussian chain) observed in Fig.8; this increase is more visible under the form an intensity increase in the straight region of log-log plot of Fig. 9. Thus it agrees with a plateau height varying as $1/l_p$.

**TABLES :**

|  | $M_0$ (g/mol) | Vmol (cm$^3$) | Mw (g/mol) | Mw/Mn | $N_w$ | Degree of sulfonation f |
|---|---|---|---|---|---|---|
| h- PS | 104 | 98 | 67 500 | 1.03 | 625 | 0 |
| d-PS | 112 | 98 | 73 000 | 1.04 | 652 | 0 |
| h- PSSNa | 206 | 108 | 150 000 | 1.12 | 730 | 1.00±0.02 |
| d-PSSNa | 213 | 108 | 170 000 | 1.2 | 800 | 0.98±0.03 |

*Table 1: Characteristics of the polymers before synthesis (from Polymer Standard Services; non deuterated polystyrene h-P and deuterated d-PS), and after Vink sulfonation, under the form of sodium salt.*

|         |   | MwPS        | Mw/Mn         | Nw                    |
|---------|---|-------------|---------------|-----------------------|
| d-PSS   |   | 73000       | 1.04          | 652                   |
| h-PSS   |   | 67500       | 1.03          | 625                   |
| Sample  |   | M0 (g/mol)  | Vmol (cm3)    | Mw estimated (g/mol)  |
| f = 0.36 | D | 148.4      | 101           | 107 000               |
|         | H | 140.7       | 101           | 101 000               |
| f = 0.64 | D | 176.6      | 105           | 127 000               |
|         | H | 169.3       | 105           | 122 000               |

Table 2: Characteristics of partially sulfonated PSSNa chains. $M_w$ is estimated from the SANS measurements, taking an average $N_w = 720$ for $N_w$, see Table 4 below.

|       |                                    | PSSNa f=1 | SSNa f=0.64 | SSNa f=0.36 |
|-------|------------------------------------|-----------|-------------|-------------|
|       |                                    | 115       | 109.5       | 103.5       |
|       | $|k|_{ZAC}$ (×10$^{-12}$ cm)       | 3.65      | 3.82        | 3.97        |
| Water | $x_{ZAC}$                          | 0.71      | 0.69        | 0.68        |
|       | $k_{S_T}$ (×10$^{-12}$ cm)         | -7.46     | -7.79       | -7.82       |
| Na$^+$ | $|k|_{ZAC}$ (×10$^{-12}$ cm)      | 0.206     | 0.211       | 0.213       |
|       | $k_{S_T}$ (×10$^{-12}$ cm)         | 0.13      | 0.13        | 0.13        |
| Br$^-$ | $|k|_{ZAC}$ (×10$^{-12}$ cm)      | 0.679     | 0.636       | 0.624       |
|       | $k_{S_T}$ (×10$^{-12}$ cm)         | -1.31     | -1.31       | -1.31       |

Table 3: Values of the different contrast lengths of polyions in solution (calculated from V. Sears, Neutron News **1992** 3 26.) for the different solvents without taking into account sodium ions. $x_{ZAC}$ is the volume fraction of deuterated solvent in the sample, $|k|_{ZAC}$ the contrast length of the units in ZAC solvent and $k_{S_T}$ the contrast length of the non deuterated units (h_PSS) in the labeled solvent (D$_2$O with or without salt). To show that we can neglect them, we give evaluations of Na$^+$ and Br$^-$ ions contrast lengths in water using the ionic radii respectively R=0.95Å et R=1.95Å taken from R. A Robinson R.H. Stokes, Electrolyte solutions, 2$^{nd}$ Ed. p 461. This is discussed again in References [48, 49].

|         | f=1             | F=0.64     | f=0.36      |
|---------|-----------------|------------|-------------|
| c=0.34M | Nw = 830 (760)  | Nw = 750   | Nw = 1130   |

|          | f=1      | F=0.64   | f=0.36  |
|----------|----------|----------|---------|
| c=0.34M  | 176±5 Å  | 97±5 Å   | 76±3 Å  |
| c=0.17M  | 197±5 Å  | 105±5 Å  | 72±3 Å  |

*Table 4: Measured apparent degree of polymerization $N_w$ and radii of gyration $R_g$ (Å) for the different samples.*

|  | $(R_g/R_{coll})^3$ | | |
| --- | --- | --- | --- |
|  | f =1 | f =0.64 | f =0.36 |
| c=0.34M | 680 | 115 | 55 |
| c=0.17M | 955 | 145 | 45 |

*Table 5: Ratios of volumes occupied by the polyions over the volume of a completely collapsed chain. $R_g$ is the radius of gyration measured and $R_{coll}$ is the radius of gyration of a chain collapsed in a state of maximum density, $R_{gcoll} \sim 20\text{Å}$ (Eq. 6).*

|  | $c_s = 0$ M | $c_s = 0.34$ M | $c_s = 0.68$ M |
| --- | --- | --- | --- |
| $R_g$ | 97 ± 5 Å | 73 ± 8 Å | 66 ± 5 Å |

*Table 6: Effect of added salt on the radius of gyration of the chains for f=0.64 at a concentration $c_p$=0.34M.*

| All sizes in Å | $c_p=0.34$M | | | $c_p=0.17$M | | |
|---|---|---|---|---|---|---|
| | $R_g$ | $R_{SD}$ $R_{Gaus}$ | $R_{overlap}$ | $R_g$ | $R_{SD}$ $R_{Gaus}$ | $R_{overlap}$ |
| f=1 | 176±5 | 227 216 | 89 | 112±5 | 254 241 | 112 |
| f=0.64 | 97±5 | 125 119 | | 105±5 | 136 129 | |
| f=0.36 | 76±3 | 98 93 | | 72 ±3 | 93 88 | |

*Table 7: Comparison of the radius $R_{overlap}$ of the volume (assumed spherical) available per chain at concentration $c_p$ (see text Eq.8 ) with: (i) the measured $R_g$ (ii)the radius of the chains calculated from the measured $R_g$ assuming two extreme conformations : a sphere $R_{SD}=(5/3)^{1/2}R_g$ , or a Gaussian chain $R_{Gauss}= (6/4)^{1/2}R_g$. One sees that these different polyion sizes are clearly inferior to $R_{overlap}$ for f=0.36 only.*

| | $c_p = 0.34M$ | | | | $c_p = 0.17M$ | | | |
|---|---|---|---|---|---|---|---|---|
| | $q_{maxS}$ (Å$^{-1}$) | $q^*$ (Å$^{-1}$) | $2\pi/q_{maxS}$ (Å) | $2\pi/q^*$ (Å) | $q_{maxS}$ (Å$^{-1}$) | $q^*$ (Å$^{-1}$) | $2\pi/q_{maxS}$ (Å) | $2\pi/q^*$ (Å) |
| f=1 | 0.115 ±0.010 | 0.106 ±0.005 | 55 | 60 | 0.090 ±0.005 | 0.078 ±0.005 | 70 | 80 |
| f=0.64 | 0.088 ±0.008 | 0.078 ±0.005 | 70 | 80 | 0.068 ±0.008 | 0.059 ±0.002 | 90 | 105 |
| f=0.36 | 0.060 ±0.008 | 0.043 ±0.002 | 105 | 145 | 0.048 ±0.005 | 0.035 ±0.002 | 130 | 180 |

*Table 8: Abscissa $q_{maxS}$ of maxima of the apparent structure factors $S_{app}(q) = S_T(q)/S_1(q)$ and $q^*$ for the peak of total scattering $S_T(q)$. There is always a slight shift. Note that peaks of $S_2(q)$ are at the same abscissa than the one for $S_{app}(q)$.*

# FIGURES

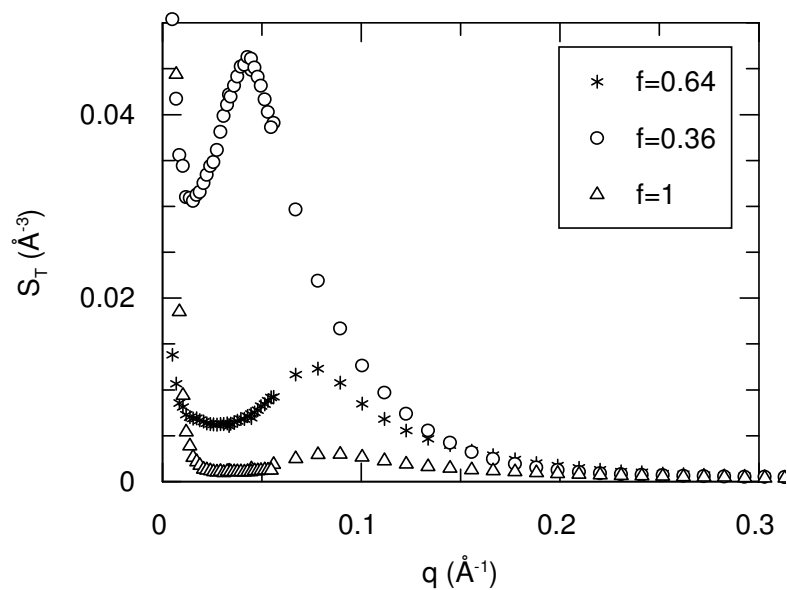

*Fig. 1: Total scattering $S_T(q)$ (all chains non deuterated in deuterated water) for three degrees of sulfonation of PSSNa in water, at a polymer concentration $c_p$=0.34M.*

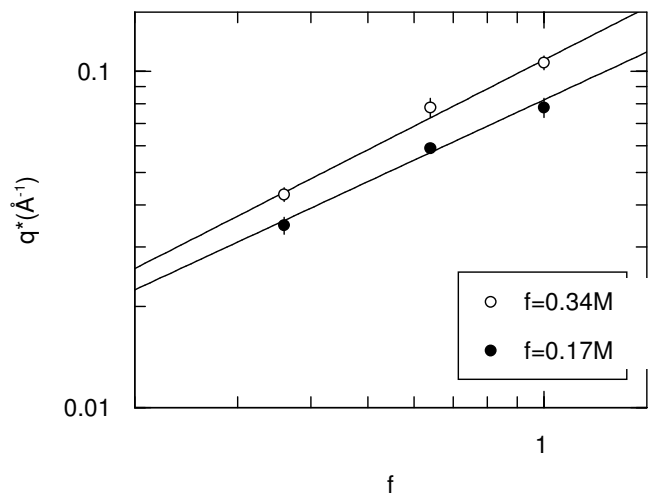

*Fig. 2: Log-log plot of q\*, the peak abscissa of $S_T$ as a function of the polyion chemical charge fraction f (sulfonation rate) for two polymer concentration $c_p$ = 0.34 and 0.17M.*

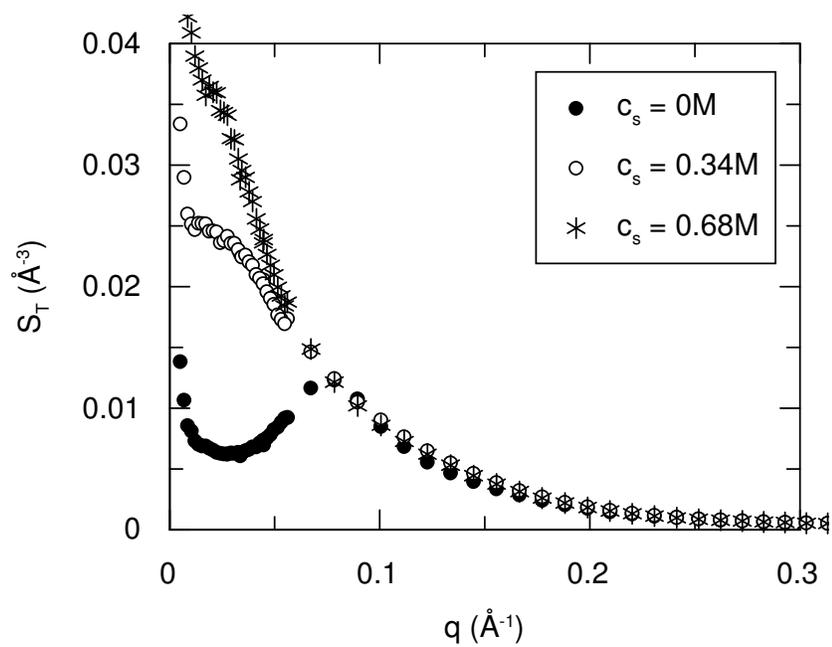

*Fig. 3: Effect of added salt concentration ($c_S$ =0, 0.34 and 0.68M) on the total scattering $S_T(q)$ of sample f = 0.64.*

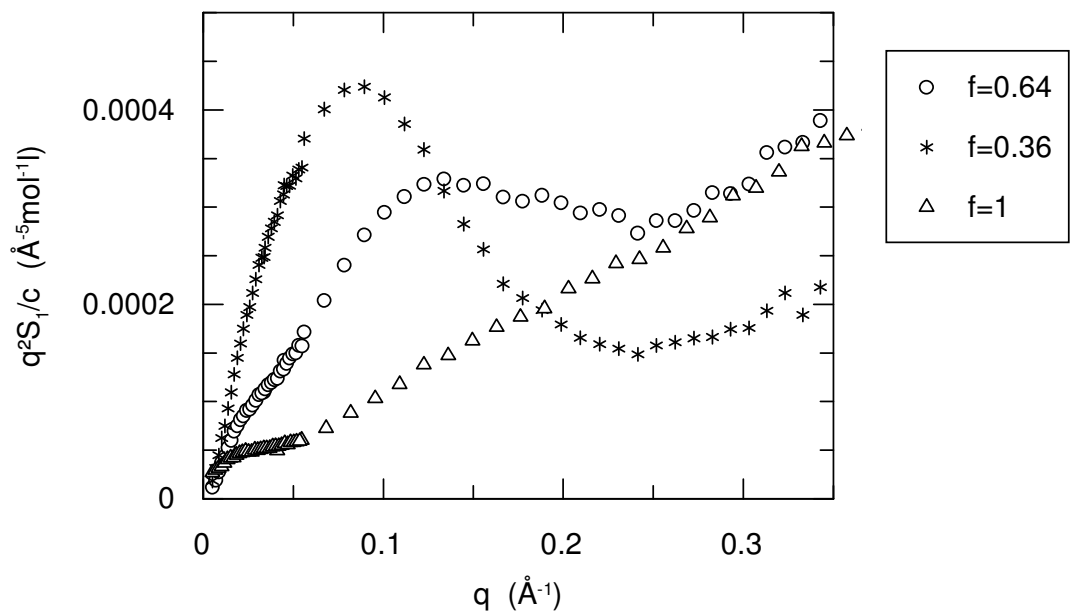

*Fig. 4: Kratky plot $q^2S_1(q)$ of the intrachain scattering function $S_1(q)$ of polyions, measured at polymer concentration $c_p=0.34M$, for different degree of sulfonation f=0, 017 and 0.34M.*

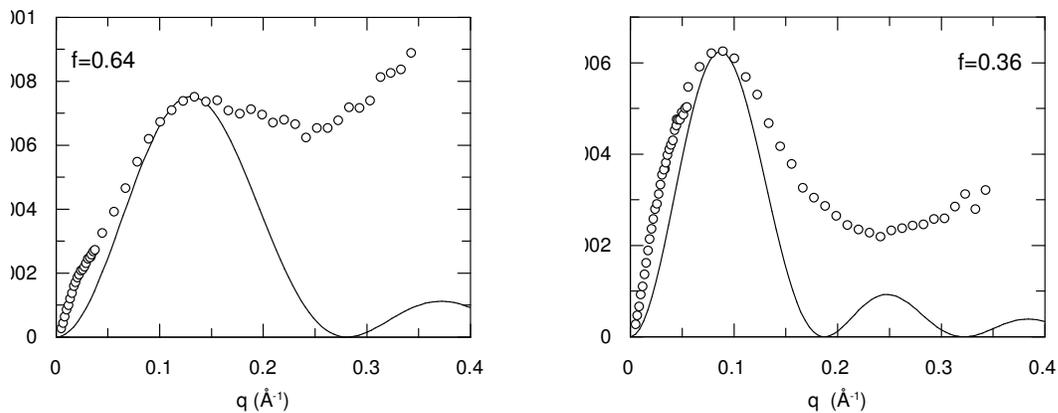

*Fig. 5: Fit in the q region where $q^2 P(q)$ is maximum, of the form factor $P(q)$ measured at polymer concentration $c_p$ is 0.34 M, by a function proportional to the form factor of a sphere, $P(q) = A.P_{Sphere}(q)$. For f=0.64, and $N_w$=730, we find R=16±1Å, and A=0.1125. For f=0.36, $N_w$ =1130, R=24±1Å and A=0.21.*

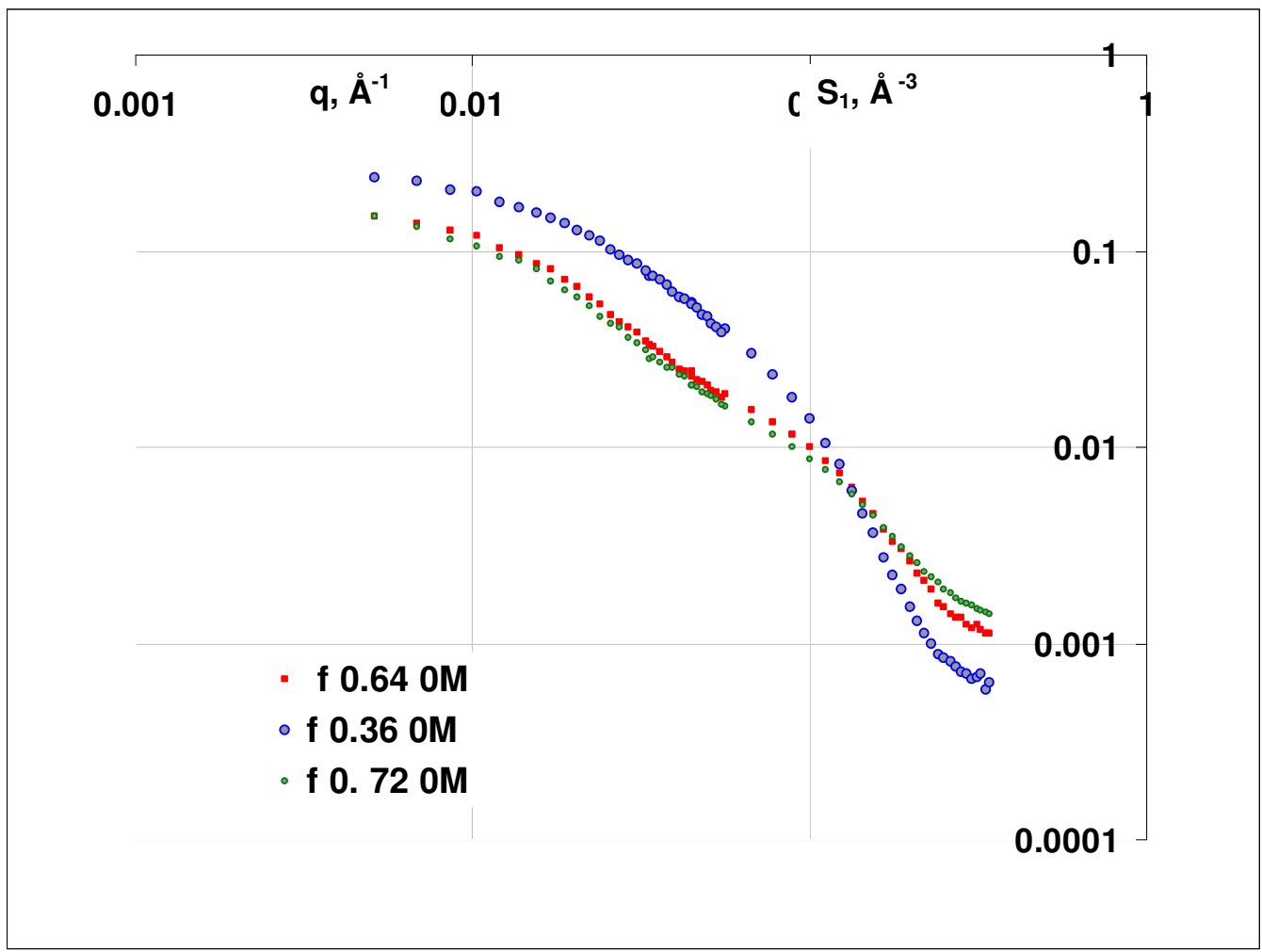

*Fig. 6: Log-log plots of the intrascattering function $S_1(q)$ for polyions of different degree of sulfonation, f=0.72, 0.64 and 0.34, at polymer concentration $c_p$=0.34M, with no salt added (0M). Data are the same shown in $q^2 S_1(q)$ plot in Fig. 4.*

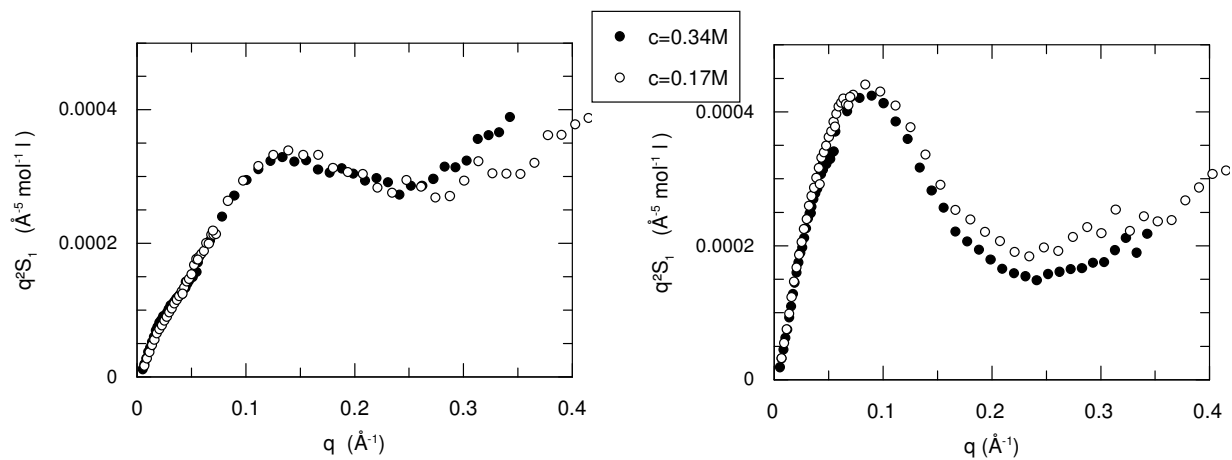

*Fig. 7: Effect of concentration in polyions $c_p$ (noted c in the figure) = 0.17 and 0.64M on the intrachain signal in Kratky plot $q^2S_1(q)$ for partially sulfonated polyions. Left hand side: f = 0.64 ; right hand side : f=0.36.*

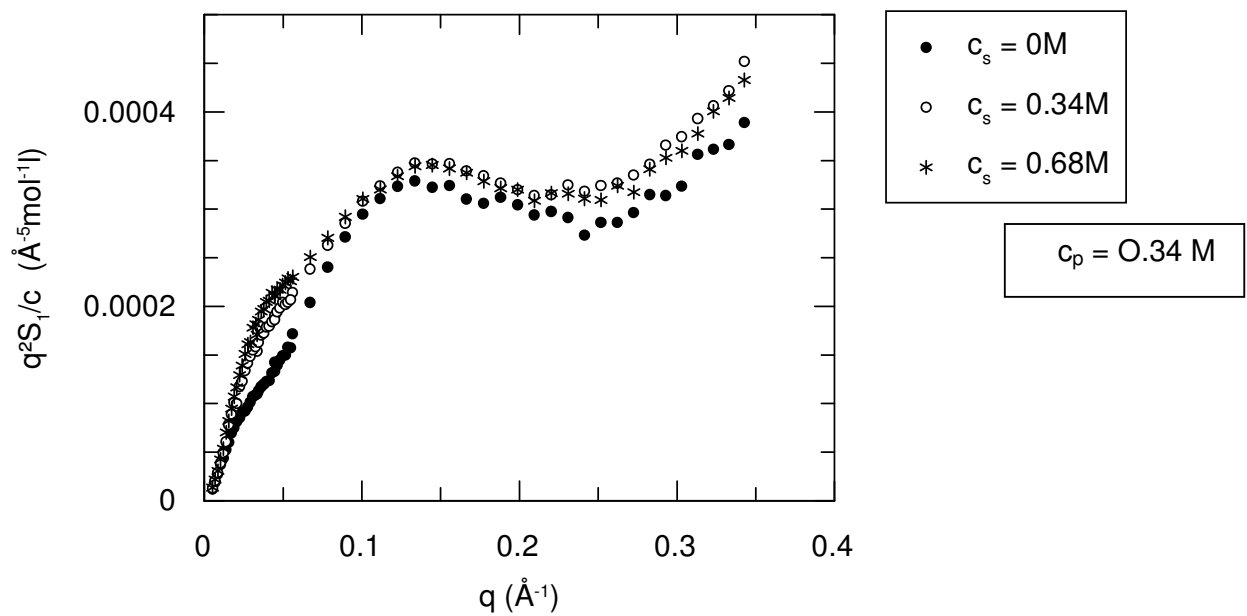

*Fig. 8: Effect of added salt concentration ($c_s$ =0, 0.34 and 0.68M) on the intrachain scattering $S_1(q)$ factor for sulfonation rate f=0.64 and $c_p$ = 0.34M*

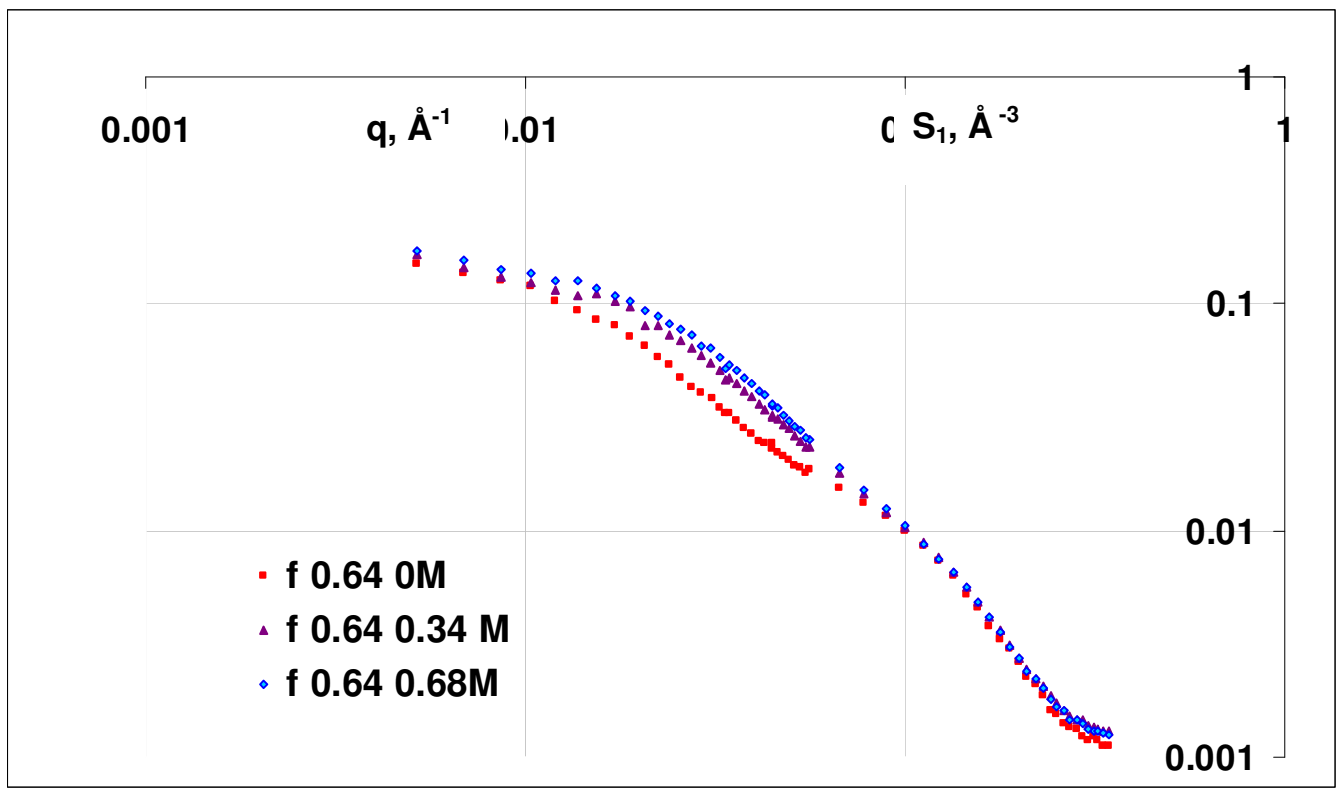

*Fig.9: Log-log plots of the intrascattering function S1(q) for polyions of degree of sulfonation f = 0.64, at polymer concentration $c_p$=0.34M, at added salt concentration 0M, 0.34M, and 0.68M.*

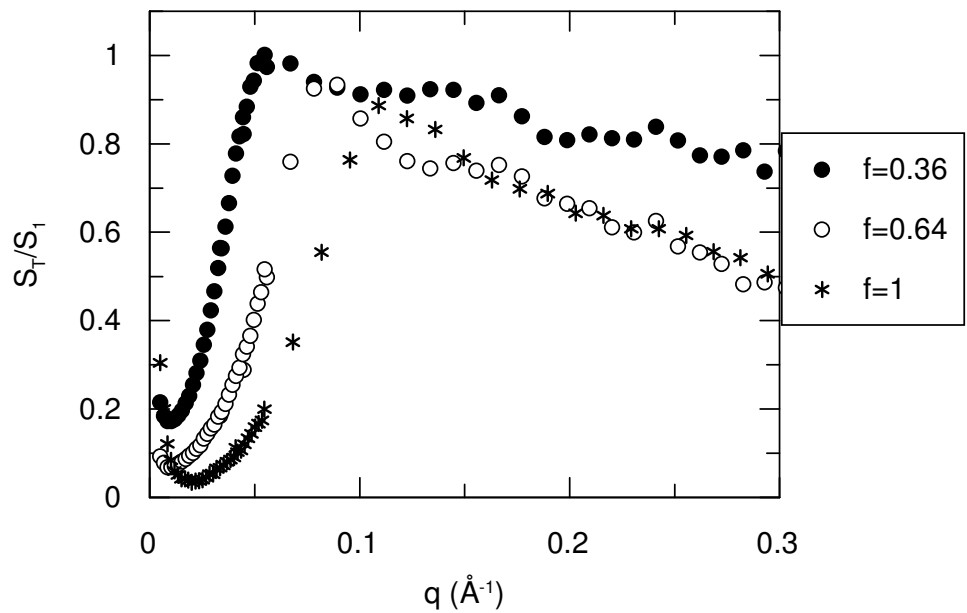

*Fig. 10: Apparent structure factor $S_{app}(q) = S_T(q)/S_1(q)$ for the three sulfonation rates $f = 1$, 0.64 and 0.36 at a polymer concentration $c_p = 0.34$ M.*

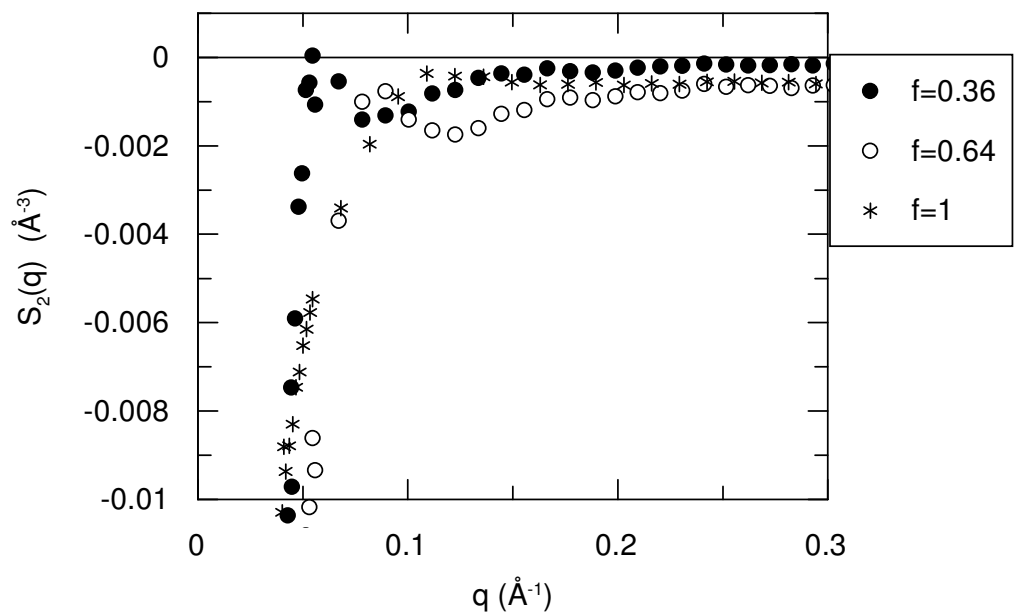

*Fig. 11: Distinct interchain scattering $S_2(q)$ for the three sulfonation rates $f= 1, 0.64$ and $0.36$, at a polymer concentration $c_p=0.34M$.*

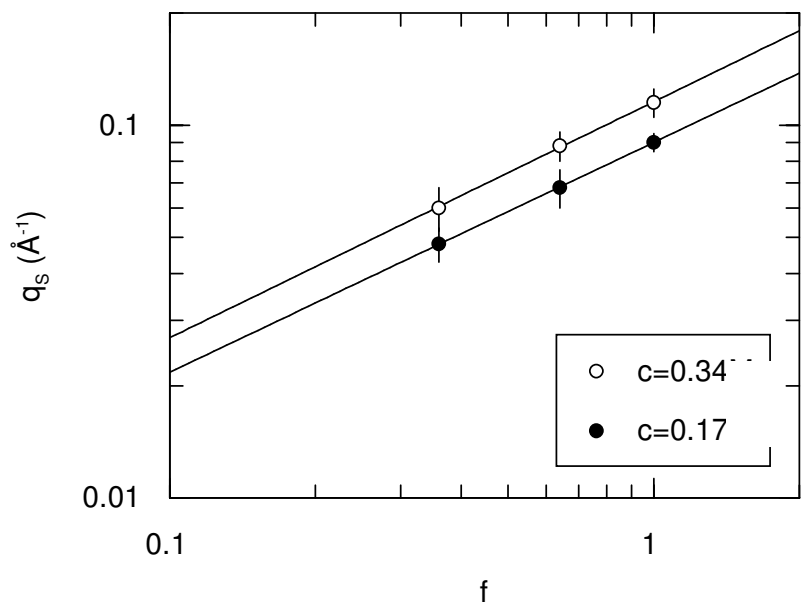

*Fig. 12: Log-log plot of the variation of $q_s$ (noted $q_{maxS}$ in the text), the peak abscissa of the apparent structure factor (see Fig. 10), versus the degree of sulfonation $f = 1$, 0.64 and 0.36, for two polymer concentration $c_p = 0.17M$ (filled circles, below) and 0.34M (open circles, above). The lines are fits of the power law $q_s \sim C^{0.63 +/- 0.003}$.*

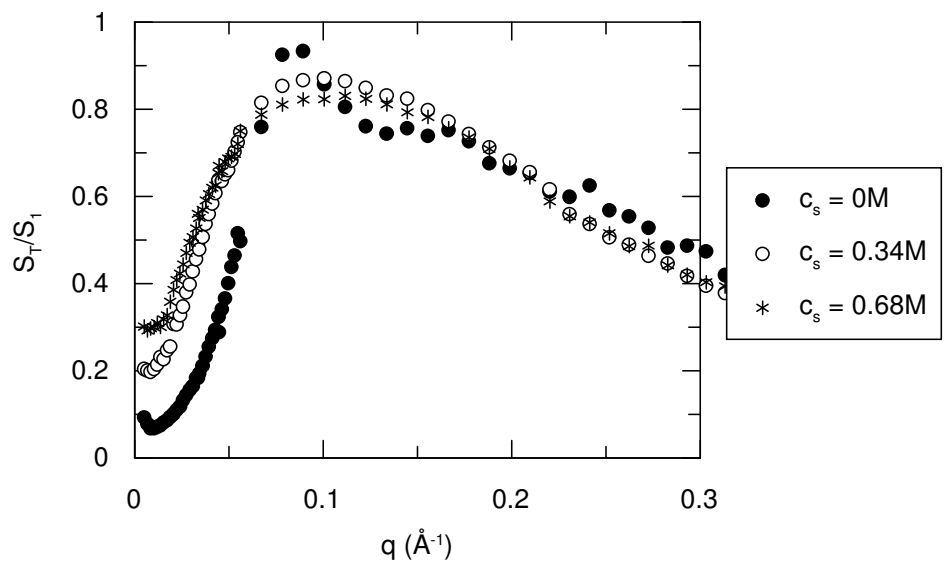

*Fig. 13: Apparent structure factor $S(q) = S_T(q)/S_1(q)$ for rate of sulfonation $f=0.64$ in presence of added salt (NaBr) at concentration $c_s = 0$, 0.34 and 0.68 M.*

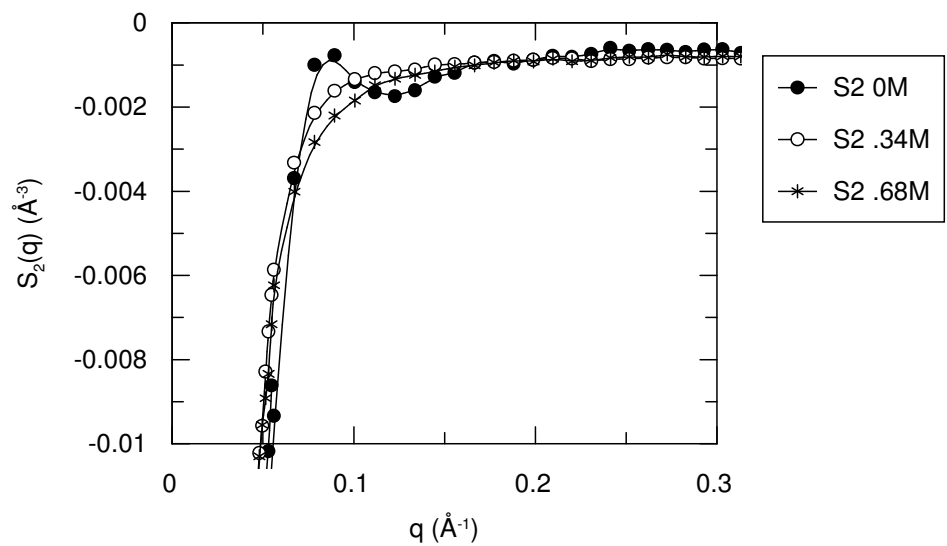

*Fig. 14: Distinct interchain scattering $S_2(q)$ for sample of degree of sulfonation f=0.64 in presence of added salt at salt concentration $c_s$ = 0, 0.34 and 0.68 M.*

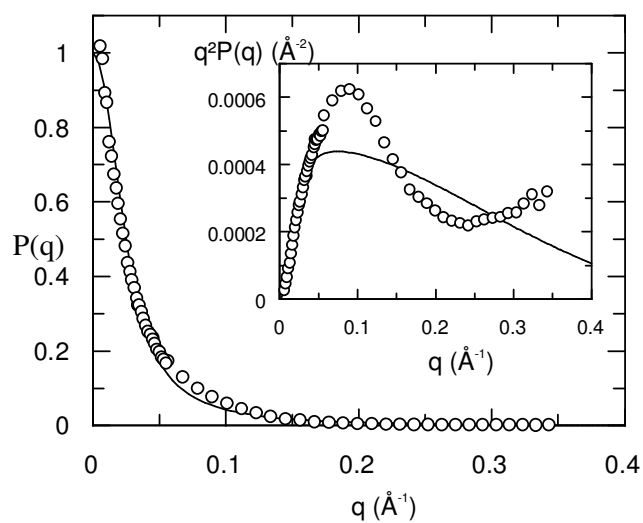

*Fig. 15: Unsuccessful attempt of fitting the measured form factor P(q) for f=0.36 at a concentration $c_p$=0.34 M by a Gaussian chain of $N_p$=149 adjacent pearls of diameter D=12.9Å. Main picture: linear – linear plot. Insert: Kratky plot.*

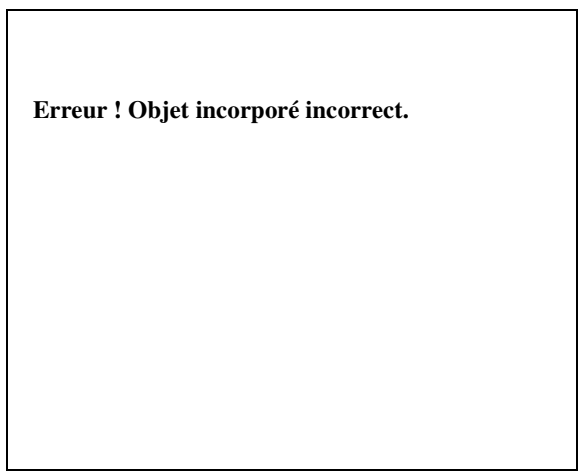 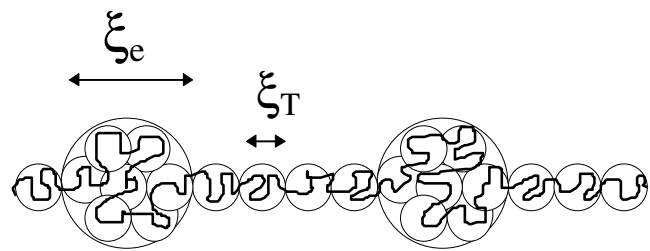

*Fig. 16: Picture for the conformation and the arrangement of chains of PSSNa partially charged in water (left), and of the Pearl necklace model (right)*

## Appendix: Comparaison with calculations.

**Comparison with Schweins-Huber analytical calculations and a Limbach-Holm simulation.**
The Schweins calculation [63] introduces a well defined distance A between the pearls of radius R. Distance A is responsible on the log-log plots for a maximum, and R for a second oscillation. A typical curve is shown in Appendix for A= 40Å and R= 7.5 Å. Depending on the mass ratio pearls over strings, these maxima are more or less pronounced, and below a ratio of typically ½ they appear like smooth oscillations which we could call shoulders. Also if A and R are close the two oscillations may not be as distinguishable. If we take as an example the case of the 64% sample (0 M), we observe a clear shoulder around 0.1 Å$^{-1}$ ; knowing that the maximum abscissa is at 0.2 Å$^{-1}$ for A=40 Å, this gives A ~ 80 Å for our data. We also show a plot of data from simulations by Limbach and Holm [34] (reported by Schweins et al in Fig. 6 of Ref. [63], and fitted by them with a trimpbel (3 pearls) model, with A = 40 Å and R=6.5 Å. There the oscillations are more dumped, as observed when less monomers are in the pearls. Other data from this group do not show such oscillations. In practice, our experimental data neither show such oscillations, unless we attribute the last kink to a second oscillation. Since it should be characteristic of the pearl size, and occurs around 0.35 Å$^{-1}$, it would correspond to very small pearls (5 Å).
In summary, there is no strong agreement with the existence of a constant distance between pearls.

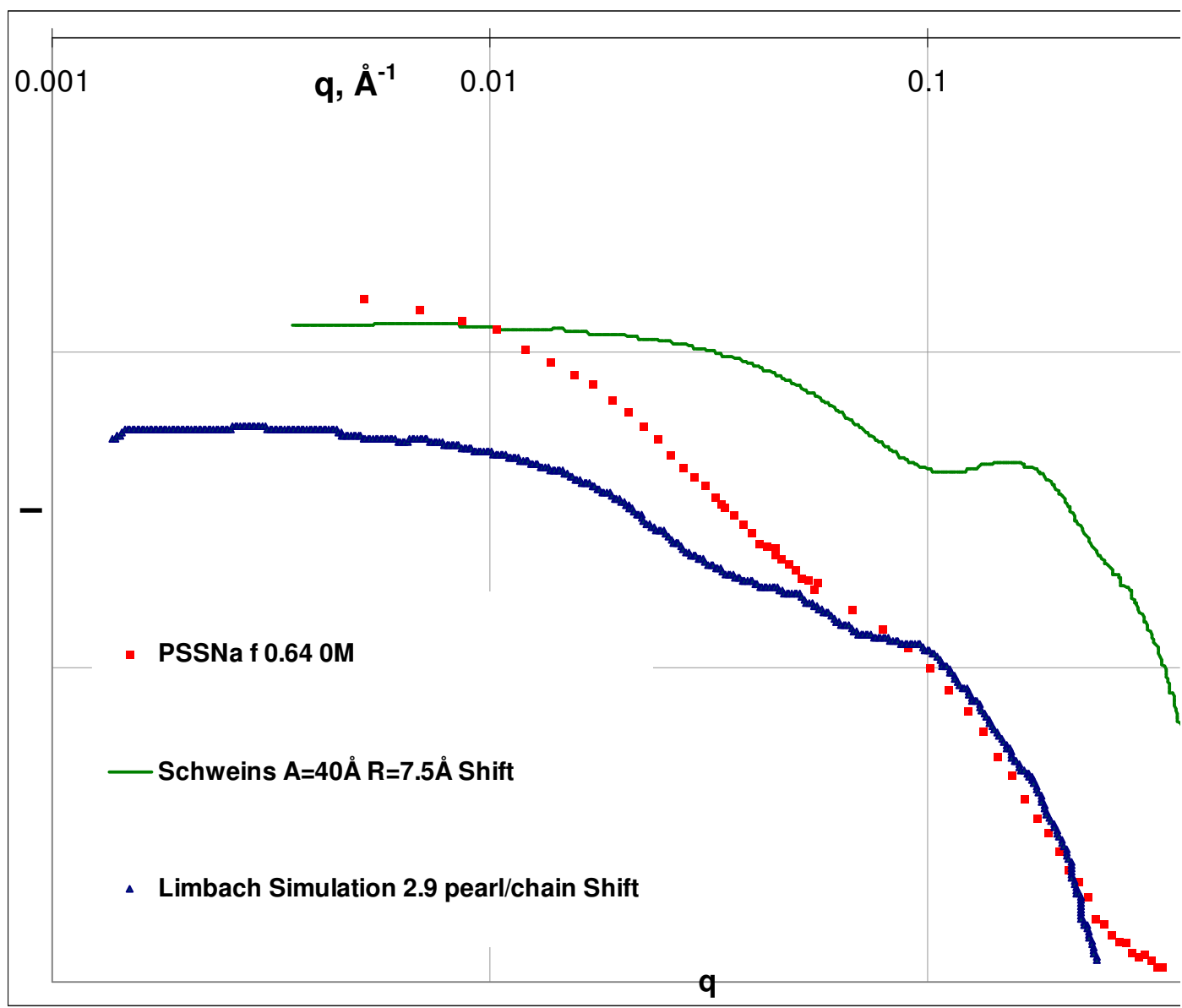

*Fig. A 1: Comparison of experimental $S_1(q)$ data for sulfonation rate $f = 0.64$ with analytical calculation of the form factor by Schweins – Huber [63] (above) for an inter – pearl distance A =*

*40Å and a pearl radius R = 7.5 Å, and with numerical simulation by Limbach – Holm [34] (data source for both curves is Fig. 6 of Ref. 63).*

**Comparison with Liao-Dobrynin-Rubinstein simulations.**
Contrary to Schweins et al., Liao, Dobrynin, and Rubinstein [64] do not obtain a constant distance between pearls, due in their opinion to couplings between concentration fluctuations and pearl formation. As a consequence, their simulated plots are much closer to our data. This was actually already remarked by the authors who published already in [64] a fit of our data for sulfonation rate f = 0.36, with 0 M salt. Here we show a temptative comparison for our other value of f, 0. 64, at 0 M salt. Data on top are genuine (not shifted) values of P(q) obtained by simulations of Ref. [64] for different values of concentration measured in $\sigma^{-3}$ unit, $\sigma$ being a unit length. When c increases, the pearl size increases, and the global radius of gyration decreases. Since scattering vector q is measured in $\sigma^{-1}$ unit, fitting with a shift on X axis is consistent with the simulation. The shift in Y axis is just adjusting the front value of $S_1(q)$ in $Å^{-3}$. When $c_p$ increases, the pearl size increases. Our best "resemblance" is for $c_p = 5 \cdot 10^{-4} \sigma^{-3}$ (good fit at large q, but $R_g$ is too small in the Guinier regime) and $c_p = 5 \cdot 10^{-5} \sigma^{-3}$. These concentrations do not really correspond to the concentration values used in our experiments; we use these plots only as a set of curves corresponding to different combinations (pearl size/global size). Though good fitting is not obtained, the available data show that a more achieved fit would be possible. At small q the deviations show that the $q^{-1}$ law (predicted by simulations at low $c_p$) is not observed, and rather is replaced by a power law of higher exponent. This is not surprising since in practice, we are in much more concentrated regime than considered in the simulations, and the chain extension is screened at this scale.

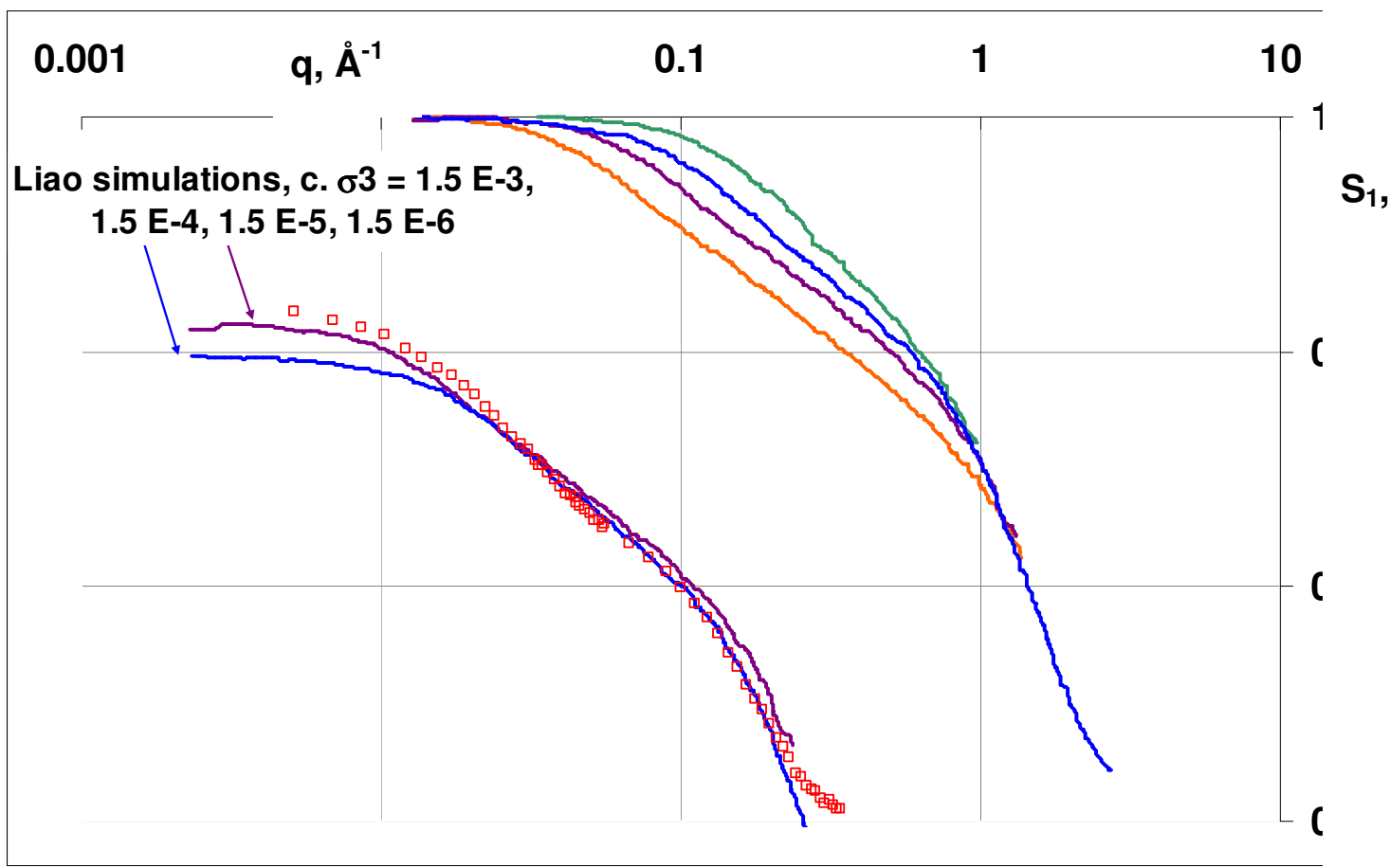

*Fig. A.2 : Comparisons with simulations by Liao et al. (data taken from [64]); above genuine (unshifted) values for $c_p$ =1.5 $10^{-3}$ $\sigma^{-3}$, 1.5 $10^{-4}$ $\sigma^{-3}$, 1.5 $10^{-5}$ $\sigma^{-3}$, and 1.5 $10^{-6}$ $\sigma^{-3}$. Fit with $c_p$= 1.5 $10^{-3}$ $\sigma^{-3}$ (above; shifted by 0.18 along X and 0.13 along Y), and 1.5 $10^{-3}$ $\sigma^{-3}$ (above; shifted by 0.18 along X and 0.09 along Y).*

**TABLE OF CONTENT GRAPHIC :**

*Fig. 1: Total scattering $S_T(q)$ (all chains non deuterated in deuterated water) for three degrees of sulfonation of PSSNa in water, at a polymer concentration $c_p$=0.34M.*

*Fig. 2: Log-log plot of q\*, the peak abscissa of $S_T$ as a function of the polyion chemical charge fraction f (sulfonation rate) for two polymer concentration $c_p$ = 0.34 and 0.17M.*

*Fig. 3: Effect of added salt concentration ($c_S$ =0, 0.34 and 0.68M) on the total scattering $S_T(q)$ of sample f = 0.64.*

*Fig. 4: Kratky plot $q^2S_1(q)$ of the intrachain scattering function $S_1(q)$ of polyions, measured at polymer concentration $c_p$=0.34M, for different degree of sulfonation f = 0, 017 and 0.34M.*

*Fig. 5: Fit in the q region where $q^2P(q)$ is maximum, of the form factor P(q) measured at polymer concentration $c_p$ is 0.34 M, by a function proportional to the form factor of a sphere, P(q) =A.$P_{Sphere}$(q). For f=0.64, and Nw=730, we find R=16±1Å, and A=0.1125. For f=0.36, Nw=1130, R=24±1Å and A=0.21.*

*Fig. 6: Log-log plots of the intrascattering function $S_1(q)$ for polyions of different degree of sulfonation, f=0.72, 0.64 and 0.34, at polymer concentration $c_p$=0.34M, with no salt added (0M). Data are the same shown in $q^2S_1(q)$ plot in Fig. 4.*

*Fig. 7: Effect of concentration in polyions $c_p$ (noted c in the figure) = 0.17 and 0.34 M on the intrachain signal in Kratky plot $q^2S_1(q)$ for partially sulfonated polyions. Left hand side: f = 0.64 ; right hand side : f=0.36.*

*Fig. 8: Effect of added salt concentration ($c_s$ =0, 0.34 and 0.68M) on the intrachain scattering $S_1(q)$ factor for sulfonation rate f=0.64.*

*Fig.9: Log-log plots of the intrascattering function $S_1(q)$ for polyions of degree of sulfonation f=0.64, at polymer concentration $c_p$=0.34M, at added salt concentration 0M, 0.34M, and 0.68M.*

Fig. 10: Apparent structure factor $S(q)= S_T(q)/S_1(q)$ for the three degrees of sulfonation f at a polymer concentration $c_p$=0.34 M.

Fig. 11: Distinct interchain **q, Å$^{-1}$** $_2(q)$ for the three degrees of sulfonation f at a polymer concentration $c_p$=0.34M.

Fig. 12: Log-log plot of the variation of $q_s$ (noted $q_{maxS}$ in the text), the peak abscissa of the apparent structure factor (see Fig. 10), versus the degree of sulfonation f =1, 0.64 and 0.36, for two polymer concentration 0.17M (filled circles, below) and 0.34M (open circles, above). The lines are fits of the power law $q_s \sim c_p^{0.63+/- 0.003}$.

Fig. 13: Apparent structure factor $S(q)= S_T(q)/S_1(q)$ for rate of sulfonation f = 0.64 in presence of added salt (NaBr) at concentration $c_s$ = 0, 0.34 and 0.68 M.

Fig. 14: Distinct interchain scattering $S_2(q)$ for sample of degree of sulfonation f=0.64 in presence of added salt, at salt concentration $c_s$ = 0, 0.34 and 0.68 M.

Fig. 15: Fit of the measured form factor P(q) for f=0.36 at a concentration $c_p$=0.34 M by a Gaussian chain of Np=149 adjacent pearls of diameter D=12.9Å. Main picture: linear – linear plot. Insert: Kratky plot.

Fig. 16: Picture for the conformation and the arrangement of chains of PSSNa partially charged in water (left), and of the Pearl necklace model (right)

**Figures in Appendix:**

Fig. A 1: Comparison of experimental $S_1(q)$ data for sulfonation rate f = 0.64 with analytical calculation of the form factor by Schweins –Huber [63] (above) for an inter – pearl distance A = 40Å and a pearl radius R = 7.5 Å, and with numerical simulation by Limbach – Holm [34] (data source for both curves is Fig. 6 of Ref. 63).

Fig. A.2 : Comparisons with simulations by Liao et al. (data taken from [64]); above genuine (unshifted) values for $c_p$ =1.5 $10^{-3}$ $\sigma^{-3}$, 1.5 $10^{-4}$ $\sigma^{-3}$, 1.5 $10^{-5}$ $\sigma^{-3}$, and 1.5 $10^{-6}$ $\sigma^{-3}$. Fit with $c_p$= 1.5

*$10^{-3}$ $\sigma^{-3}$ (above; shifted by 0.18 along X and 0.13 along Y), and 1.5 $10^{-3}$ $\sigma^{-3}$(above; shifted by 0.18 along X and 0.09 along Y).*

**END OF THE MANUSCRIPT**